\documentclass[lettersize,journal]{IEEEtran}
\usepackage{amsmath,amsfonts}
\usepackage{algorithmic}
\usepackage{algorithm}
\usepackage{array}
\usepackage[caption=false,font=normalsize,labelfont=sf,textfont=sf]{subfig}
\usepackage{textcomp}
\usepackage{stfloats}
\usepackage{url}
\usepackage{verbatim}
\usepackage{graphicx}
\usepackage{cite}
\usepackage{todonotes}
\usepackage{url}

\begin{document}

\title{Modeling and Physics-Enhanced Fault Detection in Wastewater Pump Stations}

\author{Katayoun Eshkofti, Henrik Sandberg \IEEEmembership{Fellow, IEEE}, Mikael Nilsson, and Matthieu Barreau
\thanks{K. Eshkofti, H. Sandberg, and M. Barreau are with the Division of Decision and Control Systems, Digital Futures, KTH Royal Institute of Technology, SE-100 44 Stockholm, Sweden {\tt\small \{eshkofti,hsan, barreau\}@kth.se}. M. Nilsson is with Xylem Water Solutions AB, Stockholm, Sweden (e-mail: Mikael.Nilsson@xylem.com).}}

\markboth{Journal of \LaTeX\ Class Files,~Vol.~14, No.~8, August~2021}%
{Shell \MakeLowercase{\textit{et al.}}: A Sample Article Using IEEEtran.cls for IEEE Journals}


\maketitle

\begin{abstract}
Monitoring wastewater pump stations is essential because they are critical infrastructure. However, monitoring is still often performed manually due to the lack of suitable algorithmic methods and data. This paper introduces a high-fidelity, physics-enhanced simulator of a three-pump wastewater station that captures transient hydro-mechanical dynamics at a one-second resolution. The simulator is fully parameter-driven, adaptable to other wastewater stations, and capable of generating datasets for data-driven analytics. It can also generate balanced faulty datasets when real failures are scarce or confidential. A comparison with high-frequency SCADA data from a municipal station shows strong agreement across key operational metrics. Furthermore, the paper proposes robust statistical and mathematical frameworks for fault detection and isolation, including a nested-model F-test to detect pump degradation or system faults, and a tangent residual approach to distinguish pump faults from system faults using operating-point kinematics. This framework enables what-if studies, facilitates early fault diagnosis based on flow rate and head, and provides actionable insights for condition-based maintenance in wastewater pumping infrastructure.
\end{abstract}


\section{Introduction}
\IEEEPARstart{M}unicipal wastewater pump stations are critical infrastructures in urban water management ecosystems, enabling the reliable and continuous conveyance of sewage to centralized treatment plants. Their operations directly affect environmental protection, public health safeguards, and economic efficiency. Equipment is commonly housed in enclosed structures, and because a single station typically aggregates inflows from an entire catchment area, hydraulic loads undergo significant fluctuations. This results in frequent starts and stops, surcharging, and transient events that impose high mechanical and electrical stress on pumps and valves, accelerating wear. Therefore, adopting a control and diagnostic approach enables the quantification of operating stresses, improves efficiency, extends asset life, and helps prevent overflows.

To mitigate these stresses in practice, contemporary wastewater stations employ various forms of control architecture, ranging from on/off switching to variable frequency drive (VFD)-modulated proportional-integral-derivative (PID) controllers, to optimize operation, reduce mechanical wear on pumps, and adapt to dynamic hydraulic loads. Pump stations consist of at least two pumps, where one pump is usually in standby mode to respond to failures of the main pumps. Parallel pumps are extensively deployed in wastewater applications due to high flow rate or pressure demands and broad demand variations \cite{Zhounian2020reliability}. In practice, these pumps operate in lead-lag duty cycles with rotation, where the lead pump alternates among units to equalize wear and maintain availability.

Within this operating context, centrifugal pumps are predominantly used in wastewater pump stations. These pumps are turbomachines that transport liquids by raising a specified flow volume to a particular pressure level \cite{gulich2010centrifugal}. Although pump technology has advanced significantly in recent years, degradation mechanisms such as impeller blockage, bearing wear, and internal circulation, along with system-level faults like pipe clogging, continue to pose operational challenges. Therefore, integrating fault detection mechanisms into the control system is highly beneficial \cite{jones2006pumping}. The practical issue is to develop diagnostics that distinguish pump-side faults from system-side faults.

However, detecting and identifying faults is very challenging, as they often exhibit overlapping symptoms, including reduced flow rates, increased energy consumption, and frequent start-stop cycling. This ambiguity hampers root-cause diagnosis using supervisory control and data acquisition (SCADA) alarms, which typically rely on threshold violations or power disconnection (e.g., motor protection tripping), without distinguishing between pump-related and system-related faults. Consequently, maintenance actions are often heuristic, potentially leading to unnecessary and costly replacements.

Two maintenance paradigms respond to this issue. \emph{Preventive maintenance} is considered a potential solution for improving the scheduling of maintenance operations. It is also applicable to wastewater centrifugal pumps, particularly for critical and expensive components \cite{Bianchini2019}, where maintenance intervals are based on the relationship between runtime and pump failure. In contrast, \emph{condition-based monitoring} involves real-time tracking of pump performance and degradation to reduce redundant maintenance actions and minimize costs \cite{beebe2004condition}.

Despite this need for discriminative diagnostics, most studies have primarily focused on issues related to wastewater treatment plants and drinking water distribution networks \cite{Wang2006adaptive,WaterLedesma2024,Guo2023Pump}. Research concerning pumps has mostly concentrated on design, construction, and operational characteristics \cite{gulich2010centrifugal,stoffel2015assessing,totten2011handbook}, as well as on scheduling \cite{fritzson2011simulation, zhou2024digital} and optimization \cite{DataJOHNSON,GeneticOLSZEWSKI}.

Addressing this gap through experimental investigation of pump behavior is difficult, as it is prohibitively expensive due to sensor installation and testbed requirements.
Compounding these constraints, failures are rare, costly to stage, sometimes not directly observable, may involve critical outputs that are not directly measurable, and are often confidential. Therefore, simulation serves as a dual solution and has become an essential tool. First, it allows observability and controllability of model variables, and by synthesizing labeled, class-balanced faulty datasets, it enables the development and benchmarking of diagnostics while preserving SCADA confidentiality \cite{fritzson2011simulation}. Second, simulations can mitigate the effects of real-world noise.


Even with simulation, another important aspect must be considered. To the best of our knowledge, integrated digital twins capable of simulating coupled hydro-electro-mechanical dynamics, interdependent control logic, and progressive fault propagation remain underexplored. Furthermore, while data-driven surrogates can capture transient behaviors, they often violate physical conservation laws when extrapolated beyond training data \cite{Karpatne2017, karniadakis2021physics}.

To address these limitations, a physics-enhanced wastewater pump station simulator with a statistical and mathematical fault discrimination framework has been developed. The main contributions are as follows:

\begin{itemize}
    \item A high-fidelity simulator is presented that replicates transient dynamics at 1-second resolution, incorporating practices such as soft start and stop, and supervisory sequencing logic. Affinity law-scaled pump curves and a system curve are integrated, along with power computations and sensor noise injection, to model physics-constrained hydraulic-electrical relationships.
    \item The simulator is calibrated using SCADA data from a municipal pump station, showing high accuracy in predicting real-world pump behavior. It enables the generation of balanced datasets for training machine learning algorithms while preserving SCADA confidentiality.
    \item The fully parameter-driven architecture of the simulator can be leveraged for deployment at wastewater pump stations of any size or layout, even when different types and numbers of pumps are used, provided that the site-specific geometric and technical specifications are available.
    \item A complementary two-part fault-origin isolation approach based on operating point kinematics is introduced. These methods employ only routine SCADA measurements: flow rate, head, and drive frequency. A nested-model F-test and a tangent residual index can attribute operating-point drift to either pump or system changes at low instrumentation cost. Together, they enable online detection that supports maintenance dispatch and spare-parts planning for pump stations.
\end{itemize}

This paper is organized as follows: Section~\ref{sec:configuration} develops the physics-enhanced simulator of the three-pump wastewater station under nominal and faulty conditions and outlines the control logic and operating-point kinematics. Section~\ref{sec:validation} describes the numerical implementation, validates the simulator against SCADA data of the municipal pump station, and introduces two representative fault scenarios: internal pump blockage and pipe clogging. Section~\ref{sec:dectection} presents the fault-origin isolation methods, including the nested-model F-test and the tangent residual index, which employ flow, head, and frequency measurements. Section~\ref{sec:numerical} reports numerical results on simulated datasets and compares the diagnostic approaches, and Section~\ref{sec:conclusion} concludes with key findings and suggestions for future studies.
\section{Nominal and faulty modeling}
 \label{sec:configuration}
The case study considers a municipal wastewater pump station with a cylindrical sump, where the water level $L(t)$ evolves dynamically based on mass conservation. The sump receives inflow and outflow and is equipped with water-level sensors. Pump stations may use submersible or dry-installed centrifugal pumps, typically arranged in parallel on a common discharge header. This study focuses on the submersible setup, which removes the need for a dry well and costly intake piping, reduces excavation and ventilation, enables quick pump removal and wash-down via guide bars and quick connects, and requires minimal routine attention \cite{jones2006pumping}.

\begin{figure}
    \centering    \includegraphics[width=2.5in]{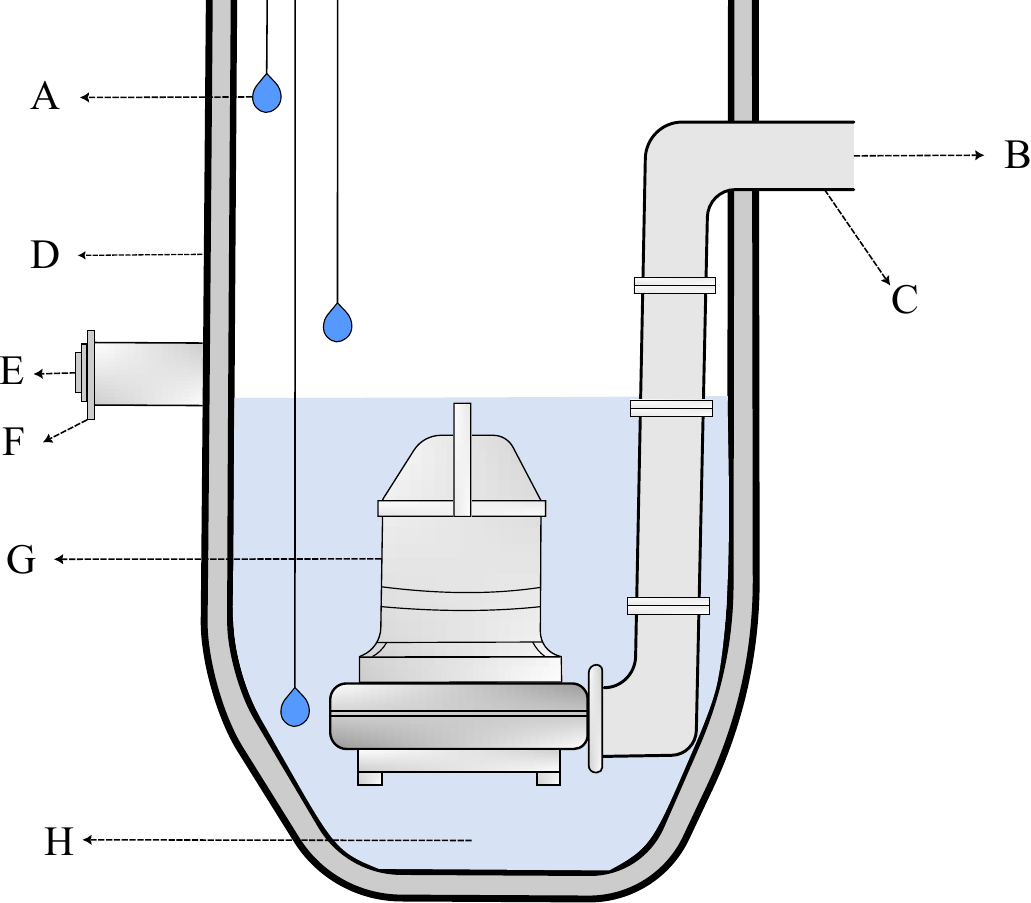}
    \caption{The typical configuration of a submersible wastewater pump station inspired from \cite{FlygtBrochure2024}. The wastewater pump station consists of (A) float switches, (B) outflow, (C) possible position of a flow rate sensor, (D) sump, (E) inflow, (F) possible position of a flow rate sensor, (G) pump, and (H) possible position of a pressure sensor.}
    \vspace{-4mm}
    \label{fig:sump}
\end{figure}

The station also includes a controller that operates the pumps based on the sump water level. The schematic of the sump and pump configuration is shown in Fig.~\ref{fig:sump}.
Level sensors typically generate analog signals to represent water level or flow rate. Float switches are installed at different sump heights and trigger when the water reaches a set value, producing digital signals. The controller’s output can switch the pump on or off, alternating between two states.

Some pumps also include VFDs, enabling the controller to adjust pump frequency based on water level. In practice, VFDs are mainly used for soft start and stop. The controller sets the motor speed to a defined value, with shaft speed ramping smoothly up or down instead of direct line energizing. This reduces electrical inrush, mechanical shock, and hydraulic pressure spikes common in traditional on/off logic.

Section~\ref{sec:configuration} has two main subsections: the nominal model and the faulty model. Subsection~\ref{subsec:nominal} details hydraulic, power, and control modeling of pumps under normal conditions, while Subsection~\ref{subsec:faulty} describes two common pump station failures and their impact on pump and station metrics.

\subsection{Nominal model} \label{subsec:nominal}
The nominal model defines the healthy baseline of pump operation and serves as the reference for subsequent analyses. Hydraulic modeling is performed by coupling each pump curve with the system curve to determine the operating point. The system curve represents the head required by the network at a given flow rate. It combines the static head with friction and minor losses resulting from the piping, bends, and check or isolation valves. Flow and head are then related to hydraulic and electrical power through an efficiency factor that varies with operating conditions. Variable-speed operation is represented via the affinity laws, while station supervisory logic and operational constraints are explicitly integrated. The following sections formalize these components and present the equations used in the simulation.

\subsubsection{Hydraulic modeling}
\label{subsubsec:hydraulic_modeling}
This section presents key terminology, equations, and characteristic curves for evaluating pump performance and simulating behavior.

\paragraph{Pump Curve}
The pump flow rate, also referred to as capacity or discharge, is denoted by $Q(t)$ and represents the pumped volume per unit time. It is usually expressed in $\text{m}^3/\text{s}$ or $\text{L}/\text{s}$ for large pumps, and in $\text{m}^3/\text{h}$ for smaller pumps.

The head, $H(t)$, is the elevation of a free water surface relative to a reference datum. Pump head is the pressure difference between discharge and suction sides, measured in meters. A centrifugal pump is typically represented by a one-dimensional curve that shows the head developed at various flow rates, assuming a constant speed and fixed impeller diameter\cite{jones2006pumping}.

Manufacturers obtain pump performance data by testing with clean, cold water. For simulation, these discrete data points are fitted with a smooth polynomial function:
\begin{equation}
H_{\text{pump}}(Q) = a_0 - a_1 Q - a_2 Q^2,
\label{eq:pumpcurve}
\end{equation}
where $H_{\text{pump}}$ signifies the pump head in meters, and $a_i$ for $i \in \{0, 1, 2\}$
are the coefficients specific to the pump design. The shut-off head $a_0$ is the maximum differential head the pump can generate when its discharge is completely closed and the flow rate is zero.

\begin{figure}
    \centering    
    \includegraphics[width=0.55\columnwidth]{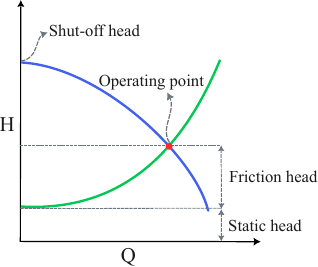}
    \caption{Operating point is the intersection of pump curve (blue) and system curve (green).}
    \label{fig:operating_point}
    \vspace{-4mm}
\end{figure}

Extended operation at or near the shut-off head heats the liquid and accelerates pump wear. The blue curve in Fig.~\ref{fig:operating_point} illustrates a typical pump curve. According to Fig.~\ref{fig:operating_point}, in order to raise the fluid to a higher point, the flow rate decreases.

\paragraph{System curve}
The resistance of pipelines introduces friction through the following relation:
\begin{equation}
H_{\text{system}}(Q) = H_{\text{static}} + kQ^2.
\label{eq:systemcurve}
\end{equation}

This equation relates system head $H_{\text{system}}$ to steady flow $Q$, defining the system curve with two components: static head and friction head. The static head $H_{\text{static}}$ is the elevation the liquid must overcome, independent of flow rate, and sets the vertical offset of the friction head loss curve. The friction head $kQ^2$ represents the head required to overcome losses caused by friction in the piping, valves, and fittings of the system where the pump operates, is the loss coefficient. This coefficient, obtained from the Darcy--Weisbach equation {\cite[Chapter~3]{jones2006pumping}}, depends on pipe geometry (length and diameter), flow rate, friction factor, and gravity.

Closing valves, reducing pipe diameter, or clogging increase head loss. The green curve in Fig.~\ref{fig:operating_point} depicts the system curve, which shifts upward as frictional head increases. The shut-off head in \eqref{eq:pumpcurve} must exceed the static head; otherwise, the pump cannot overcome the initial height and no flow develops. Unlike the pump curve, which depends on pump design and characteristics, the system curve is installation-specific and shifts with changes in liquid levels, pipework, or control elements.

\paragraph{Operating point}
Once the pump is installed in a given system, the head developed by the pump must equal the dynamic head loss in the system. In other words, the intersection of the pump curve and the system curve yields a single point known as the operating point shown in Fig.~\ref{fig:operating_point}.

Pipe fouling from sludge increases losses and steepens the system curve. Conversely, pump aging and wear on bearings, impeller, and other parts shift the pump curve downward and left. Both effects reduce flow rate and move the operating point left, while the vertical shift depends on which effect dominates.
\subsubsection{Power} \label{subsubsec:power}
Pump output power refers to the work performed per unit of time to lift water to a higher elevation and is formulated as follows:
\[
P_{\text{output}} = \rho g Q H,
\]
where \( P_{\text{output}} \) is the output power in kW and \( \rho \), \( g \), and \( H \) denote the fluid density (\(\text{kg}/\text{m}^3\)), gravitational acceleration (\(\text{m}/\text{s}^2\)), and total head (m), respectively. The pump input power is computed as:
\[
P_{\text{input}} = \frac{\rho g Q H_{\text{pump}}}{\eta_p(Q,f)}.
\]

Here, \( P_{\text{input}} \) and \( \eta_p \) denote the input power and pump efficiency, respectively. Pump efficiency determines the portion of motor input power used to move the water and it strongly varies with speed $f$ and flow. 

\subsubsection{Affinity laws}
\label{subsubsec:affinity_laws}
Key metrics for analyzing pump performance under varying conditions include energy transfer, flow, head, and power. The affinity laws define the governing rules of pump performance with respect to changes in speed. If the performance metrics at a nominal speed are known, their values at other speeds can be inferred using the following affinity laws:
\[
\frac{Q_{\text{nominal}}}{Q_1} = \frac{f_{\text{nominal}}}{f_1}, \quad  
\frac{H_{\text{nominal}}}{H_1} = \left( \frac{f_{\text{nominal}}}{f_1} \right)^2,
\]

\[
\frac{P_{\text{nominal}}}{P_1} = \left( \frac{f_{\text{nominal}}}{f_1} \right)^3,
\]
where \( f_{\text{nominal}} \) is the nominal frequency of the pump in Hz, and \( Q_{\text{nominal}}, H_{\text{nominal}}, \) and \( P_{\text{nominal}} \) are the corresponding nominal flow rate, head, and power, respectively.

Tests on various centrifugal pumps show that the affinity laws accurately estimate flow rate and head but are less reliable for power, as power depends on efficiency, which changes with speed and operating point.


%

\subsubsection{Control logic and operational constraints}
\label{subsubsec:control_logic}
In wastewater pump stations equipped with three pumps using a VFD mechanism, two pumps must meet the design flow rate, while the third serves as standby. At low flow, a single lead pump operates. However, if the lead pump is active but the water level rises above a certain threshold, meaning the inflow rate exceeds the discharge rate of the lead pump, the lag pump is activated. Both pumps then run until the inflow drops below the lead pump’s discharge.

To simulate the operation of a pump station with three pumps, two approaches can be adopted. According to manu\-facturers, wastewater pumps typically employ a closed loop control strategy, where the VFD is modulated by a level feedback PID controller \cite{jones2006pumping}.

In practice, many stations run in open loop. In this configuration, the pump starts with a gradual speed increase, known as soft start, operates continuously at constant speed, and performs a soft stop when a stop condition is met. This mode is more common because inflow is usually lower than the discharge at the pump’s predefined minimum speed; thus, after soft start, no further speed variation occurs.

\paragraph{Common supervisory start and stop logic}
The control policy governing the pump duty cycle to prevent short cycling is expressed as:
\[
n(t^{+}) =
\begin{cases}
\min\{ n(t) + 1, \, N_{\max} \}, & \text{if } L(t) \geq S_{n(t)+1}, \\[6pt]
\max\{ n(t) - 1, \, 0 \}, & \text{if } L(t) \leq E_{n(t)}, \\[6pt]
n(t), & \text{otherwise},
\end{cases}
\]
where $N_{\max}$ is the maximum number of pumps installed in the pump station, \( L(t) \) is the measured noisy water level in the sump, and \( n(t) \in \{0, 1, 2, N_{\max}\} \) represents the number of pumps currently running. The upper threshold for each pump $i \in \{ 1, 2, N_{\max}\}$ is denoted by $S_i$ that triggers the operation of the $i^{\text{th}}$ pump, and $E_i$ represents the lower thresholds such that $E_i < S_i$.

Based on this logic, if the water level reaches the start threshold and no pump is running, one pump is activated. The lead pump is chosen by a round-robin method to distribute wear and thermal load evenly.
\paragraph{Soft start and soft stop controller}
The soft start and stop mechanisms follow a ramp profile. The frequency command \( f(t) \) sent to the VFD is time dependent and defined as:
\resizebox{\linewidth}{!}{$
f(t) =
\left\{
\begin{array}{ll}
f_{\text{min}} + r t, & 0 \leq t < t_{\text{ramp}}, \\
f_{\text{max}},       & t_{\text{ramp}} \leq t < t_{\text{stop}}, \\
f_{\text{max}} - r (t - t_{\text{stop}}), & t_{\text{stop}}, \leq t < t_{\text{stop}} + t_{\text{ramp}}, \\
f_{\text{min}},       & t_{\text{stop}} + t_{\text{ramp}} \leq t < t_{\text{stop}} + t_{\text{ramp}} + t_{\text{dwell}}, \\
0,                    & t > t_{\text{stop}} + t_{\text{ramp}} + t_{\text{dwell}},
\end{array}
\right.
$}
where $r = \frac{f_{\text{max}} - f_{\text{min}}}{t_{\text{ramp}}}$. Here, $f_{\text{min}}$ and \( f_{\text{max}} \) are the minimum and maximum frequencies of the pump, and \( t_{\text{ramp}} \) is a user defined ramp time. The pump ramps from $f_{\text{min}}$ to $f_{\text{max}}$ over $t_{\text{ramp}}$ seconds and then runs at nominal speed until a stop command is received. When the stop threshold is reached, the frequency decreases linearly to $f_{\text{min}}$, and the motor stops after an optional dwell time $t_{\text{dwell}}$.

\subsection{Faulty model}
\label{subsec:faulty}
Simulating faults is essential for analyzing wastewater pump stations under abnormal conditions. Failures are rare and costly to replicate, limiting data for diagnostics. Modeling common faults such as internal blockage or pipe clogging provides a controlled environment to study their effect on pump and station metrics. This enables balanced dataset generation, supports reliable fault detection algorithms, and facilitates condition-based and predictive maintenance strategies that reduce costs and minimize service interruptions.

\subsubsection{Pump internal blockage}
\begin{figure}
    \centering
    \includegraphics[width=0.55\columnwidth]{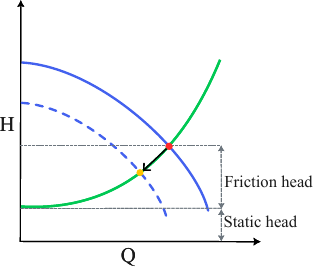}
    \caption{Operating point shift along the system curve during pump blockage.}
    \vspace{-4mm}
    \label{fig:pumpfault}
\end{figure}

When debris lodges in the pump volute or impeller passages, it disrupts flow patterns and energy transfer, while system piping and hydraulic resistance remain unchanged. Similar effects may arise from impeller wear or severe internal recirculation, though impeller wear is usually a gradual process compared to sludge buildup or sudden entry of large objects. Fig.~\ref{fig:pumpfault} depicts how the operating point shifts along the system curve when pump internal blockage occurs. This disturbance is introduced into the simulator to assess its response to such degradation.

\subsubsection{Pipe clogging}
\begin{figure}
    \centering    \includegraphics[width=0.65\columnwidth]{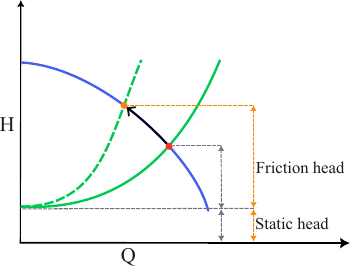}
    \caption{Operating point shift along the pump curve caused by valve throttling.}
    \vspace{-4mm}
    \label{fig:typicalsystemfault}
\end{figure}

Wastewater rising mains rarely remain clean, as progressive deposits of fats, wipes, solids, and trapped air pockets gradually raise the hydraulic resistance of the pressure main \cite{HE2013,Kargar2024}. Besides dynamic head loss from clogged pipes, static head may rise when air voids persist at high points \cite{Lubbers2006,POTHOF2011,Warda2019}. Valve throttling also increases frictional head, as shown in Fig.~\ref{fig:typicalsystemfault}. In contrast to pump faults, the system curve steepens and the operating point shifts along the pump curve.

\section{Model validation}
\label{sec:validation}
This section investigates the algorithmic details, modeling assumptions, parameters, and generated outputs of the simulator. As the goal is model validation, results are also included.

All simulations use a one-second time resolution. The time horizon is unrestricted, and any duration can be chosen. Resolution may be refined for high-resolution output or coarsened when less detail is sufficient. If high-resolution data is not necessary, coarser time steps may be used. For analyzing pump dynamics and faults, however, fine resolution provides clearer insights and patterns. A complete Python implementation will be released on GitHub upon
paper acceptance.

\subsection{Nominal condition} \label{subsec:nominalvalidation}
To guarantee the credibility of the simulator, validation begins under nominal behavior, the healthy baseline of pump station operation. This step is fundamental, as it puts forward a direct comparison between simulated results and real-world SCADA data, thereby reinforcing confidence in the accuracy of the underlying hydraulic, electrical, and control sub-models. Focusing first on nominal operation allows systematic evaluation of pump behaviors such as flow–head relationships, cycle times, and start counts. These conditions provide the benchmark for quantifying deviations when faults are introduced.
\subsubsection{Geometry and hydraulic sub-model}
The sump is modeled as a cylindrical volume with a constant cross-sectional area of \( A = 8\,\text{m}^2 \). With free surface elevation above the pump suction denoted by \( L(t) \), the sump dynamics follow the mass balance equation:
\begin{equation}
\label{eq:mass_balance}
A \frac{dL(t)}{dt} = Q_{\text{in}}(t) - \sum_{i=1}^{3} Q_i(t),    
\end{equation}
where $Q_{\text{in}}(t)$ is the inflow rate to the sump with $\Delta t = 1 \, \text{s}$ and $Q_i(t)$ represents the discharge rate from each operating pump.

The pump station is equipped with three identical pumps. The pump curve is taken from an industrial datasheet for a mid-range submersible wastewater pump with a nominal frequency of 50 Hz and a rated motor power of 15 kW \footnote{Flygt n-technology pump.\url{www.xylem.com/sv-se/products--services/pumps/packaged-pump-systems/submersible-pumps/wastewater-pumps/n-technology-pumps/}}
. The pump curve used in the simulator, scaled according to the affinity law, is expressed as:
\[
H_{\text{pump}}(Q, N) = a_0 N^2 + a_1 N Q + a_2 Q^2,
\]
where \( N = \frac{f}{f_{\text{nominal}}} \) is the scaling parameter.

The system curve combines a static head of 2 meters with a quadratic loss term, defined by a frictional coefficient of \( 3 \times 10^{-4} \, \frac{\text{s}^2}{\text{m}^5} \), and is formulated as in \eqref{eq:systemcurve}. At each simulation time step, the operating flow rate is computed by solving the intersection between the affinity-scaled pump curve and the system curve using a bisection method.

\subsubsection{Motor and power specifications}
To compute the hydraulic shaft power $P_\text{output}$, clean water density is taken as $1000 \,\text{kg}/\text{m}^3$ and gravitational acceleration $g$ as $9.81 \,\text{m}/\text{s}^2$. The head and flow rate at the operating point are injected, and an efficiency of 0.9 is assumed for the pumps, reflecting the assumption that they are new.

Electrical input power is computed using the three-phase equation \cite{fleckenstein2016}:
\[
P_{\text{input}} = \sqrt{3} \times V \times I_i \times \cos \varphi,
\]
where the phase-to-phase voltage \( V \) is set to 400 V. The nominal current \( I_{\text{nominal}} \) is 30 A, and the power factor \( \cos \varphi \) is 0.9. During ramp-up and ramp-down phases, the current is scaled proportionally to the frequency according to $I_i = I_{\text{nominal}}N,$ but is also capped at five times the nominal current to ensure that locked-rotor conditions are avoided and that inrush current does not exceed physically feasible values.

Both hydraulic and electrical power computations incorporate a Gaussian noise term with a standard deviation of 1 percent to model the sensor noise inherent in practical measurements. For instance, for the flow rate, this can be expressed as
\[
Q_{\text{measured}} = Q_{\text{true}}\cdot (1 + \epsilon), 
\quad \epsilon \sim \mathcal{N}(0, 0.01^{2}).
\]

This type of noise is based on the typical performance specifications of industrial sensors \cite{fraden2016handbook}.
\subsubsection{Soft start and soft stop}
When the water level rises to $1.6$ m with no pumps running, the lead pump starts. Conversely, when the level falls to $0.5$ m or lower, each active pump stops. These thresholds are defined by the station owners, including the City of Stockholm and Stockholm Vatten och Avfall (SVOA). The lead pump is chosen in round-robin order (Pump 1 \(\rightarrow\) Pump 2 \(\rightarrow\) Pump 3 \(\rightarrow\) Pump 1 \(\rightarrow \dots\)).  

Multi-pump operation occurs when the influent slightly exceeds the lead pump’s capacity, triggering the second pump. In the simulation, the start level for the second pump (with the lead pump already running) is $1.8$ m. When the level falls to $0.8$ m and the influent is within the lead pump’s capacity, the second pump stops.

A start command initiates a linear ramp from 0 Hz to 50 Hz over 10 seconds, crossing 25 Hz after 5 seconds. At 50 Hz, the pump runs at maximum speed. 

A stop command triggers a reverse ramp, reducing frequency linearly to zero over 10 seconds. If received during ramp-up, the command toggles the ramp direction and resets the timer to emulate the actual drive behavior.
\subsubsection{Data recording}
At one-second intervals, the algorithm records water level, inflow, aggregate outflow, pump frequency, hydraulic power, input power, and head for each pump.  

Post-processing aggregates this data into hourly energy consumption, daily start counts, daily runtimes, and cumulative energy consumption. The dataset also supports computing additional statistics for fault diagnosis and performance assessment.
\subsubsection{Field verification with sump watcher dataset}
The dataset was collected from a sump monitoring system in Stockholm, operated by SVOA and equipped with Flygt (Xylem) submersible centrifugal pumps. It captures high-frequency sensor measurements between 20 November 2024, 12:00:00 p.m. and 16 December 2024, 11:59:00 p.m. (a duration of 27 days and 14 hours). Raw data were logged by the on-site SCADA system every second and saved to a CSV file. Key variables include timestamp with a granularity of 1 s, pump operational states (0 = inactive, 1 = active) for three pumps, voltage and current per phase, water level, and power consumption. The dataset exhibits a quasi-periodic structure that reflects the cycling behavior of pumps, with voltage stability (238.4 $\pm$ 1.7 V).

A crucial preprocessing step was conducted to address systematic level measurement errors identified during 20 November 2024, 9:53 a.m. to 27 November 2024, 9:45 a.m. A constant of +0.80 m was added to the level rows falling within this duration. After correction, 1,266,952 rows (13.8\%) were modified, preserving the true water head dynamics. Additional preprocessing included duplicate removal and gap audits, which flagged gaps over 10 s for later imputation.  

\begin{figure}
    \centering
    \includegraphics[width=\columnwidth]{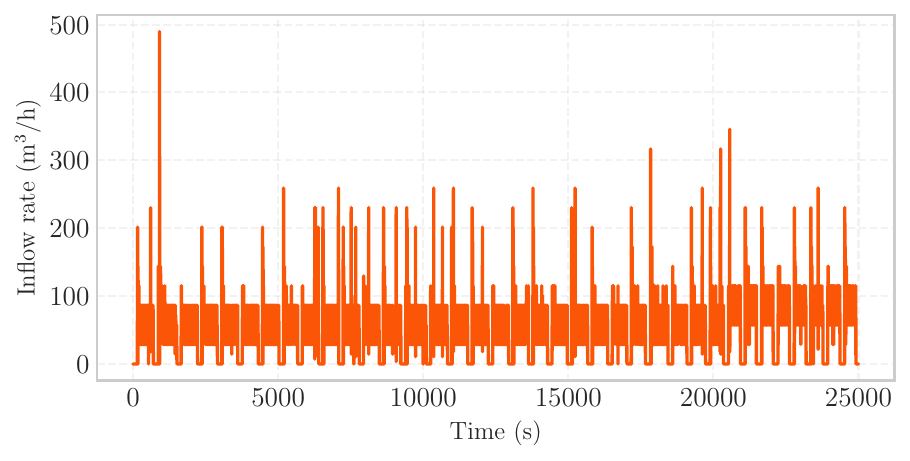}
    \caption{Observed inflow rate distribution from real-time monitoring data.}
    \label{fig:sumpinflow}
    \vspace{-5mm}
\end{figure}

The inflow rate of the wastewater sump was not provided in the dataset. Since the cross-sectional area of the sump is known and is equal to $A = 8.0 \, \text{m}^2$, the inflow rate was inferred using the centered finite difference method, as described below:
\[
Q_{\text{in}}(t_i) = A \frac{\Delta L_i}{\Delta t_i} + \sum_{j=1}^{3} z_j(t_i) Q_j(t_i),
\]
where $\Delta L_i = L_i - L_{i-1}$ and $\Delta t_i = t_i - t_{i-1}$. Forward and backward one-sided differences are used at the boundaries. Additionally, rows with $|\Delta t|$ less than $0.1 \, \text{s}$ or $|Q|$ greater than $0.4 \, \text{m}^3\text{s}^{-1}$ are discarded as duplicates or meaningless spikes. In the second term, \( z_j \in \{0,1\} \) is the state of pump \( j \). For each active pump, \( Q_j(t_i) \) can be obtained from the operating point intersection evaluated at the current wet well level \( L(t_i) \). Thus, \( A \Delta L / \Delta t \) approximates the net outflow, and adding \( \sum_j z_j Q_j \) recovers the true inflow. Moreover, negative flows are removed since the focus is on inflow dynamics. The resulting inflow series contains 2,237,754 samples, and the statistics are summarized in Table~\ref{tab:inflow}. The distribution of inflow rate for the first 25000 seconds is also depicted in Fig.~\ref{fig:sumpinflow}.

\begin{table}
\caption{Summary statistics of the inferred inflow rate.\label{tab:inflow}}
\centering
\begin{tabular}{|c||c|}
\hline
\textbf{Inflow Statistic} & \textbf{Value} \\
\hline
Median inflow & 0.016~m$^3$/s \\
\hline
95th percentile inflow & 0.032~m$^3$/s \\
\hline
99th percentile inflow & 0.040~m$^3$/s \\
\hline
\end{tabular}
\vspace{-4mm}
\end{table}

\subsubsection{Parameter calibration}
To implement the simulation, the real inflow rate is injected into the simulator. Key simulator parameters and their values are listed in Table~\ref{tab:pump_parameters}, forming the basis for the results in the next subsection.
\begin{table}
\caption{Hydraulic and operational parameters for the simulator.\label{tab:pump_parameters}}
\centering
\resizebox{\linewidth}{!}{%
\begin{tabular}{|c||c|}
\hline
\textbf{Parameter} & \textbf{Value} \\
\hline
Pump nominal frequency & 50 Hz \\
\hline
Pump nominal speed & 1460 rpm \\
\hline
Start level for the lead pump & 1.6 m \\
\hline
Stop level for all pumps & 0.5 m \\
\hline
Start level for the second pump (when one is already running) & 1.8 m \\
\hline
Stop level of the second pump (when two are running) & 0.8 m \\
\hline
Static head & 2.0 m \\
\hline
Friction coefficient & $3 \times 10^{-4}$ \\
\hline
Water density & 1000 kg/m$^3$ \\
\hline
Gravitational acceleration & 9.81 m/s$^2$ \\
\hline
Soft start / soft stop time & 10 s \\
\hline
\end{tabular}%
}
\vspace{-4mm}
\end{table}
\subsubsection{Goodness of fit analysis}
\begin{figure}
    \centering
    \includegraphics[width=\columnwidth]{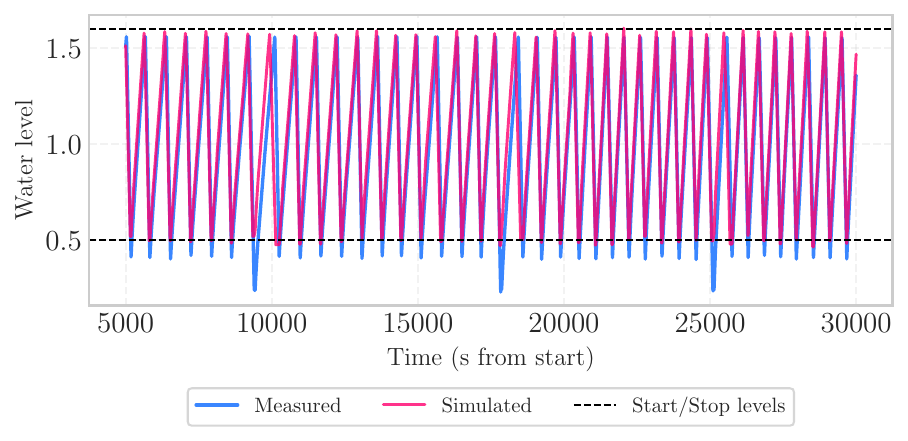}
    \caption{Validation of simulated vs.\ measured water level dynamics.}
    \vspace{-5mm}
    \label{fig:level_valid}
\end{figure}

\begin{figure}[!t]
    \centering
    \includegraphics[width=\columnwidth]{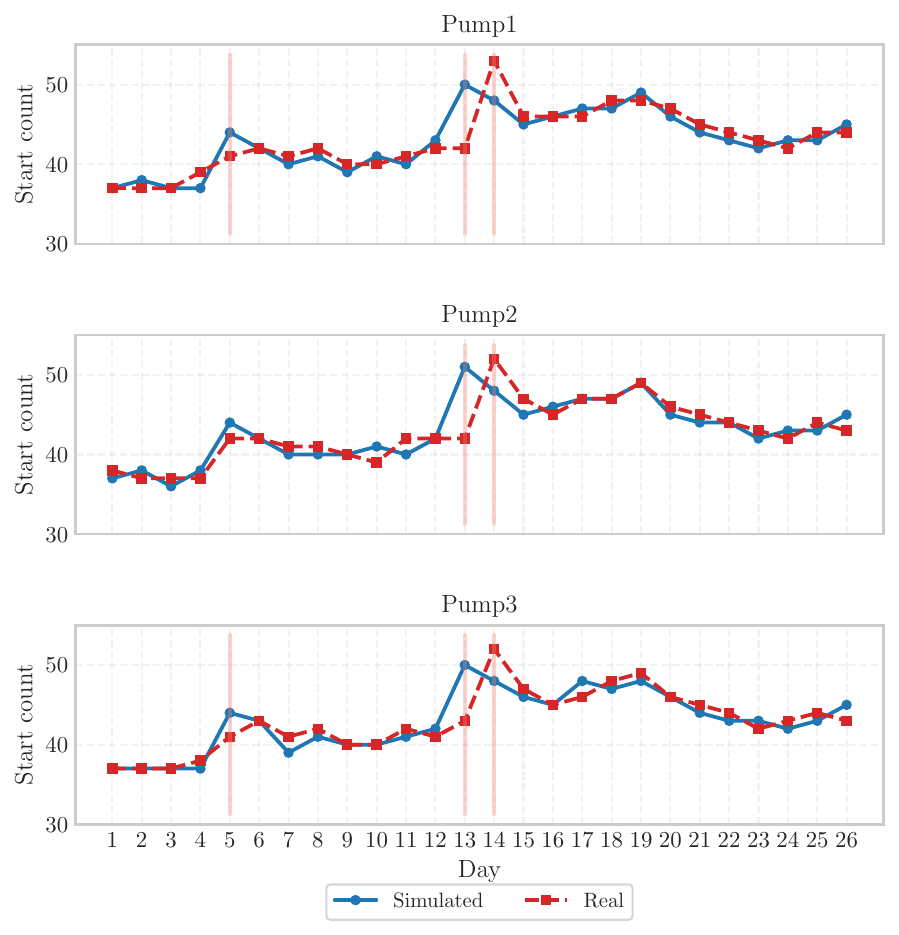}
    \caption{Comparison of simulated versus measured daily pump start counts.}
    \vspace{-4mm}
    \label{fig:start_valid}
\end{figure}

\begin{figure}
    \centering
    \includegraphics[width=0.95\columnwidth]{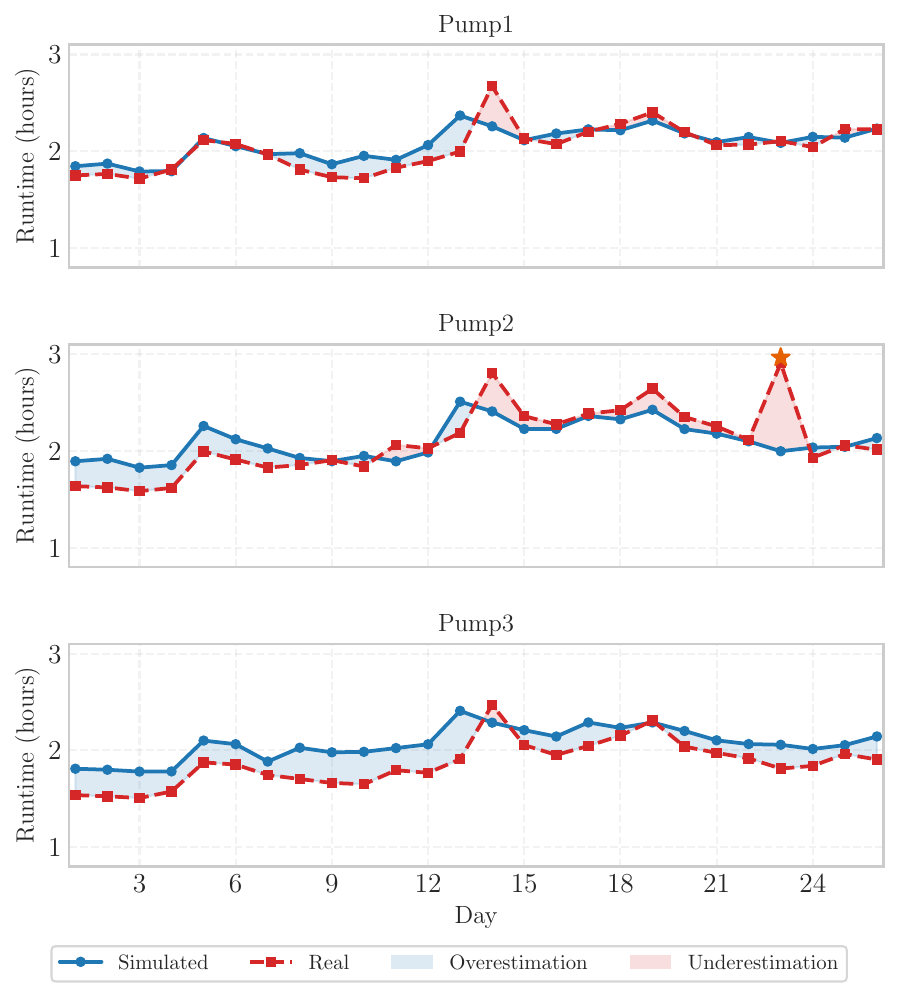}
    \caption{Comparison of simulated versus measured daily  runtimes.}
    \vspace{-4mm}
    \label{fig:runtime_valid}
\end{figure}

To validate simulator accuracy under measured inflow, three key metrics are assessed: (a) water-level dynamics, (b) pump start frequency, and (c) pump runtime. Real SCADA measurements are used for comparison. The validation spans 26 full days, and goodness of fit is evaluated by trend alignment, error distribution, and pattern consistency. Results are shown in Figs.~\ref{fig:level_valid} to ~\ref{fig:runtime_valid}.

As shown in Fig.~\ref{fig:level_valid}, the water level validation demonstrates impressive phase synchronization between the simulated and measured levels throughout the 30{,}000-second time window. In other words, the simulator accurately replicates the periodic oscillation pattern. Over the entire 26-day period, the normalized mean absolute error is 7.48\% indicating high fidelity in simulating the hydraulic response.
Fig.~\ref{fig:start_valid} demonstrates high accuracy in predicting operational cycling. Vertical highlights indicate days with deviations of three or more pump starts. Overall, all pumps show consistent accuracy. For Pump 1 and Pump 3, only three days are highlighted, while Pump 2 shows only two days with deviations exceeding three start events.

As illustrated in Fig.~\ref{fig:runtime_valid}, runtime analysis reveals occasional over- and underestimation, none significant. Deviations above 0.5 h are marked with a star and occur only once for Pump 2; all others remain below this threshold. Such deviations may stem from system losses like frictional resistance, which depend on pipe condition and related factors. Accurately determining the friction coefficient is therefore nontrivial.
\subsection{Faulty condition}
To complement the nominal condition analysis, Section~\ref{sec:validation} is extended to cover faulty operating behavior, specifically internal pump blockage and pipe clogging. These faults were selected because they are among the most common and critical, directly affecting hydraulic performance, energy consumption, and overall reliability. The following parts examine these failures and their impact on pump metrics.
\subsubsection{Pump internal blockage scenario} \label{subsub:blockagge}
The real inflow rate typically exhibits random variability and occasional surges, requiring ensembles to propagate input uncertainty to water level, test fault detection robustness under variable loading, and preserve the probability of rare high-inflow events.

\begin{figure}
    \centering
    \includegraphics[width=0.95\columnwidth]{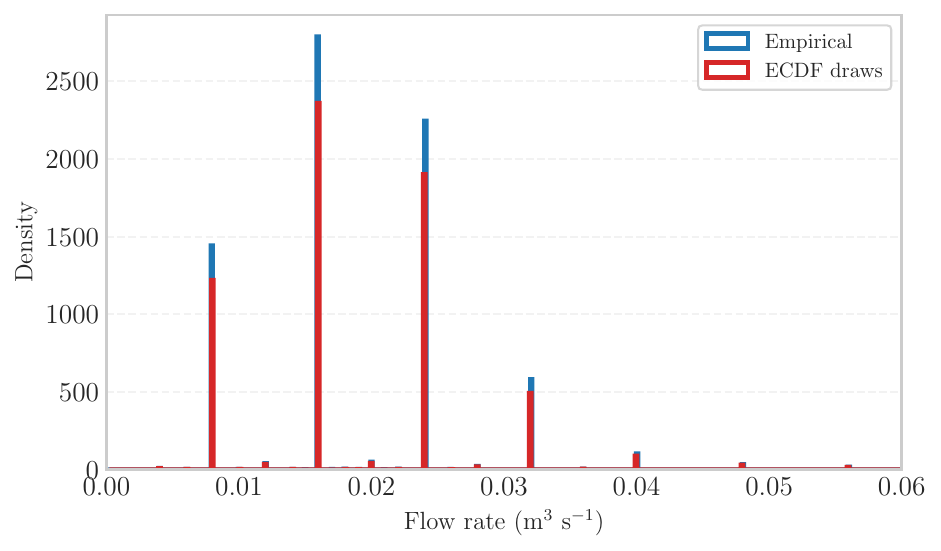}
    \caption{Density distribution (histogram) with overlaid empirical cumulative distribution function (ECDF) of flow rate measurements.}
    \vspace{-4mm}
    \label{fig:ecdf}
\end{figure}

For the nonparametric inflow sampler, let $\{q_j\}_{j=1}^m$ denote the measured inflow samples. The empirical cumulative distribution function (ECDF) sampler is defined as
\[
\hat{F}_Q(q) = \frac{1}{m} \sum_{j=1}^{m} \mathbf{1}\{q_j \leq q\}.
\]
A baseline inflow is then generated by inverse transform sampling as
\[
u_t \sim \mathcal{U}(0,1), 
\qquad 
Q_{\text{base}}(t) = \hat{F}_Q^{-1}(u_t).
\]
Here $Q_{\text{base}}$ matches the observed marginal distribution without assuming any parametric form. A comparison between the ECDF of 1,000,000 samples and the empirical data is shown in Fig.~\ref{fig:ecdf}.

According to Fig.~\ref{fig:ecdf}, the ECDF aligns closely with the data. A steep initial rise appears at very low flow rates, consistent with the sharp peak in the density plot. The curve levels off rapidly beyond 0.02~m$^3$/s, reflecting the decline in density toward zero at higher flow rates. This indicates a highly right-skewed distribution, with most flow measurements relatively low.  

To simulate transient high-inflow events, a stochastic peak model is added to the base inflow and formulated as:
\[
Q_{\text{total}}(t) = Q_{\text{base}}(t) + \sum_{i=1}^{N_{\text{peak}}(t)} Q_{\text{peak}} \cdot \mathbb{I}_{[t_i, t_i + \tau]}(t),
\]
where \( Q_{\text{total}}(t) \) and \( Q_{\text{base}}(t) \) are the total inflow rate and baseline flow rate sampled from ECDF at time \( t \), respectively. \( Q_{\text{peak}} \) denotes the magnitude of a peak event, and $N_{\text{peak}}(t)$ is the number of peak events up to time \( t \). The indicator function \( \mathbb{I}_{[t_i, t_i + \tau]}(t) \) is equal to 1 if \( t \in [t_i, t_i + \tau] \), and 0 otherwise. Moreover, \( t_i \) and \( \tau \) are the start time of the \( i^\text{th} \) peak and the duration of each peak, respectively.

In the numerical implementation, peak arrivals are generated from a homogeneous Poisson process with a rate of \( \lambda = 0.0005~\text{s}^{-1} \), corresponding to approximately 43 events per day on average. In addition, \( Q_{\text{peak}} = 30~\frac{\text{m}^3}{\text{h}} \), which is 100\% above the baseline, and \( \tau = 900~\text{s} \), equivalent to 15 minutes.

\begin{figure}
    \centering
    \includegraphics[width=\columnwidth]{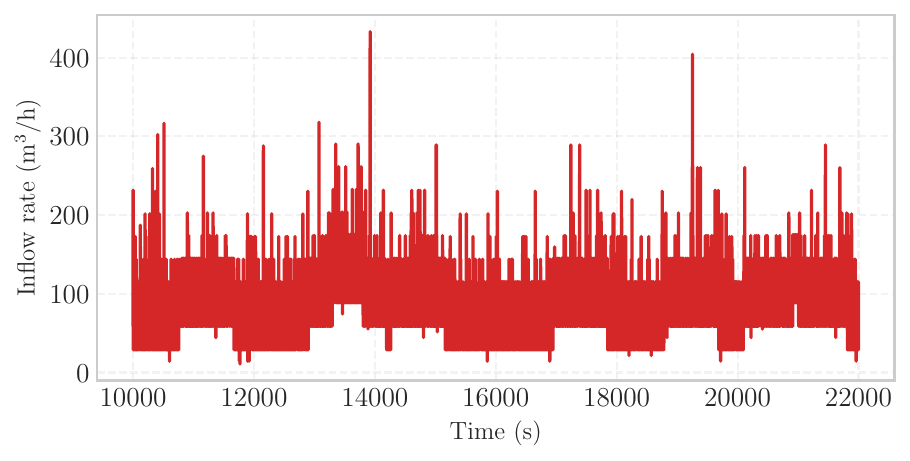}
    \caption{Inflow rate profile generated using a Poisson-modulated surge model.}
    \vspace{-4mm}
    \label{fig:inflow}
\end{figure}

These surges of wastewater can alter pump start-stop statistics, energy consumption patterns, control system response times, and may also trigger multi-pump operation. Therefore, this pump internal blockage scenario adopts the stochastic inflow model to investigate how the simulator deals with random surges of wastewater. Fig.~\ref{fig:inflow} depicts the inflow rate pattern for the given time interval.

\paragraph{Pump fault profile}
The blockage is integrated as a linear ramp beginning at $t = 12{,}000\,\text{s}$ and ending at $t = 21{,}000\,\text{s}$ at pump 1. A dimensionless blockage factor $\beta(t)$ is used to derate the impeller speed:
\begin{align*}
\beta(t) &= 1 - 0.4\, \xi(t), \\
\xi(t)   &= \frac{t - 12000}{9000}, \quad 12000 \leq t \leq 21000, \quad \xi(t) \in [0, 1].
\end{align*}

Internal blockage is realized by multiplying the commanded rotational speed, denoted by $\omega_{\text{cmd}}$ in rpm, by $\beta(t)$:

\[
\omega_{\text{effective}}(t) = \beta(t)\, \omega_{\text{cmd}}(t),
\]
where $\omega_{\text{effective}}$ is the speed that water actually experiences after partial obstruction. Therefore, when the pump is clean, \( \omega_{\text{effective}} = \omega_{\text{cmd}} \). The pump curve is updated according to the affinity laws.

\begin{figure*}[!t]
  \centering
  \includegraphics[width=0.9\textwidth]{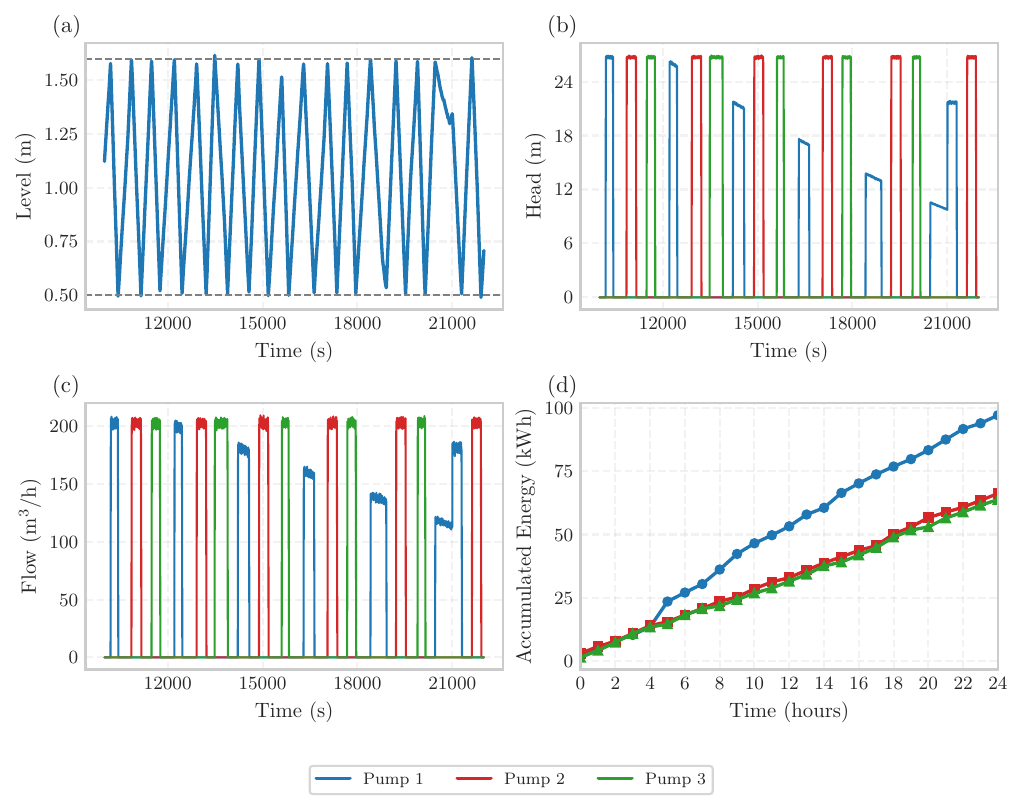} 
  \caption{Pump and station metrics: (a) sump water level, (b) pump head, (c) pump flow rate, and (d) accumulated energy, considering internal pump blockage.}
  \vspace{-4mm}
  \label{fig:metrics}
\end{figure*}

\paragraph{Pump fault simulation results}
Results are shown for a 48-hour simulation horizon. Fig.~\ref{fig:metrics} displays the distributions of water level, head, flow rate, and accumulated energy for each pump. The first three subplots capture the behavior over the time interval \( t \in [10000, 22000] \), while the last one demonstrates the given metric over a 24-hour period.

\begin{figure}
    \centering
    \includegraphics[width = \columnwidth]{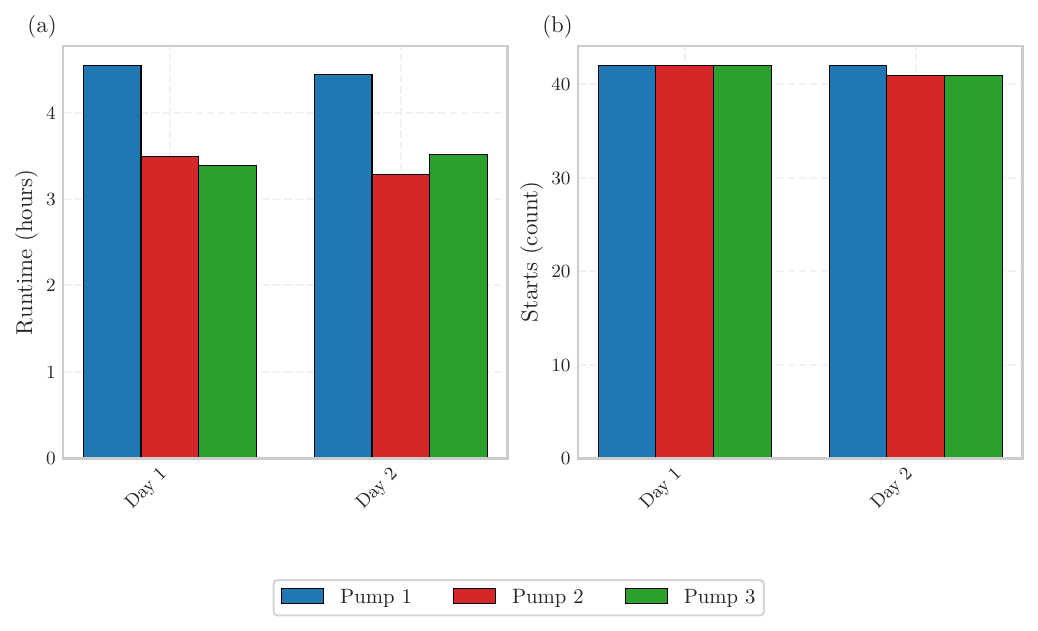}
    \caption{(a) Runtime and (b) number of starts for each pump over a two-day simulation interval, considering the internal pump blockage scenario.}
    \vspace{-4mm}
    \label{fig:runstart}
\end{figure}

As seen in Fig.~\ref{fig:metrics}(a), the water level does not exceed the upper limit during this interval and thus does not trigger multi-pump operation. Internal blockage causes reduced flow and head in pump 1, evident in Figs.~\ref{fig:metrics}(b)--(c). Fig.~\ref{fig:metrics}(d) shows accumulated energy of pump 1 rising faster than normal, as the motor draws higher input power to overcome the obstruction. The impact of internal blockage on pump runtime and start count is shown in Fig.~\ref{fig:runstart}.

\begin{figure}
    \centering
    \includegraphics[width = \columnwidth]{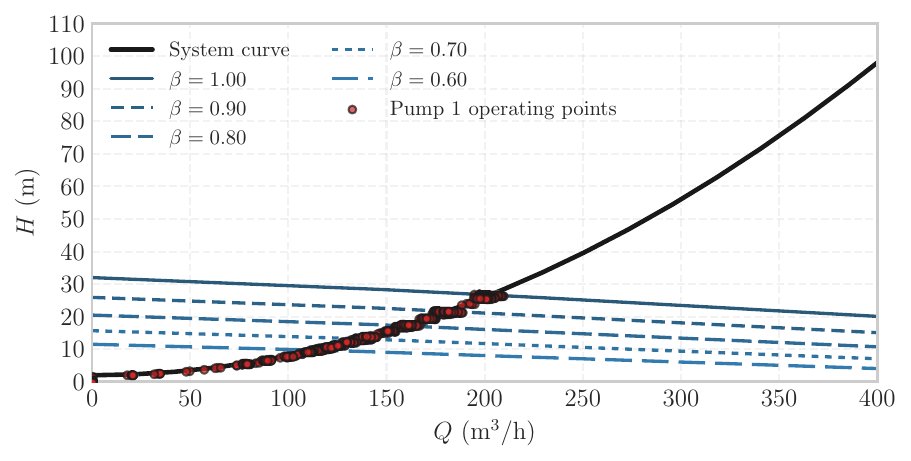}
    \caption{Operating point trajectory during the internal pump blockage fault.}
    \vspace{-4mm}
    \label{fig:trajectory}
\end{figure}

As shown in Fig.~\ref{fig:runstart}(a), blockage increases runtime, as reduced flow capacity prolongs pumping to maintain level control. In contrast, Fig.~\ref{fig:runstart}(b) shows the 2.5 h blockage has negligible effect on daily start count, since start decisions depend on level thresholds rather than pump performance. Fig.~\ref{fig:trajectory} shows the trajectory of pump~1's operating point during a simulated blockage event.

The trajectory highlights three distinct operational phases: initial unimpaired operation, progressive blockage transition, and a severe blockage state. The operating point shifts downward and left. Since the performance curve is scaled by the affinity laws, internal blockage produces a distinct hydrodynamic signature that can serve as a diagnostic feature for data-driven detection models.
\subsubsection{Pipe clogging scenario}
\label{subsub:clogging}
In the previous section, the internal pump blockage model and its effect on various metrics were discussed. This section introduces a system fault, outlining the expected hydraulic and control response and its impact on metrics.  

\begin{figure*}
    \centering
    \includegraphics[width=0.75\textwidth]{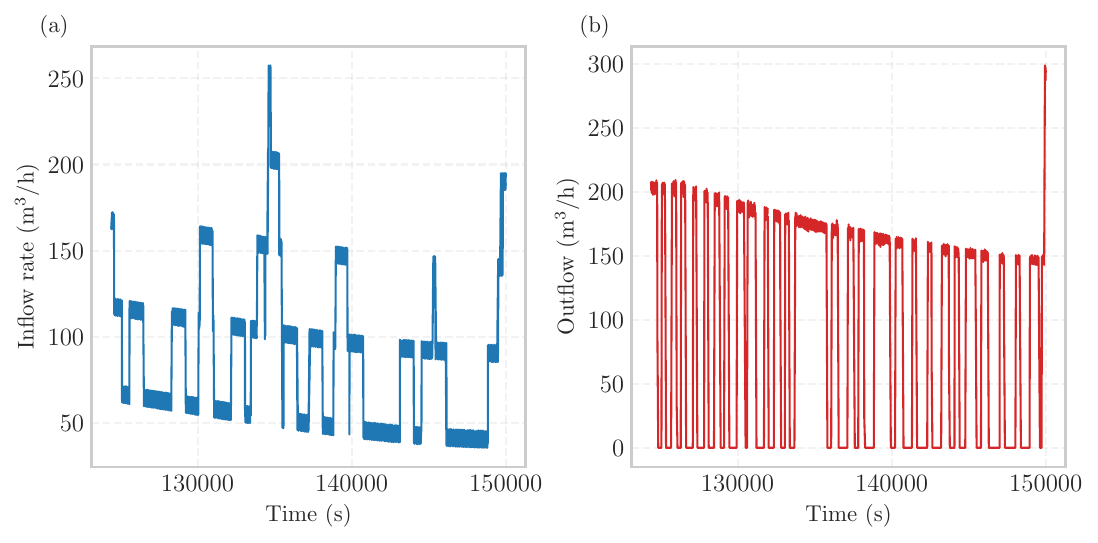}
    \caption{Distribution of (a) inflow and (b) outflow of the pump station within the given time interval.}
    \vspace{-5mm}
    \label{fig:outtflow}
\end{figure*}

Here, the inflow model combines a deterministic diurnal pattern with stochastic peaks to test the simulator under another inflow type. A sinusoidal profile oscillating between 40 and $80~\mathrm{m^{3}/h}$ over 24 h generates daily variations. As in the previous scenario, Poisson-distributed peaks are superimposed, but with bounded sensor-like noise $\varepsilon(t) \sim \mathcal{N}(0, 5^2)$ and peak magnitude $Q_\text{peak} = 50~\mathrm{m^3/h}$ to examine the multi-pump mechanism and level control strategy. Fig.~\ref{fig:outtflow} shows the inflow and outflow rates. The inflow model is:
\begin{equation*}
    \begin{array}{rl}
        Q_{\text{total}}(t) =&\!\!\!\! 60 + 20 \sin\left( \frac{2\pi t}{86400} \right) + \varepsilon(t) \hspace*{2cm}\\
        &\hfill+ \sum_{i=1}^{N_{\text{peak}}(t)} Q_{\text{peak}} \cdot \mathbb{I}_{[t_i,\, t_i + \tau]}(t).
    \end{array}
\end{equation*}

\begin{figure*}[!t]
    \centering
    \includegraphics[width=0.9\textwidth]{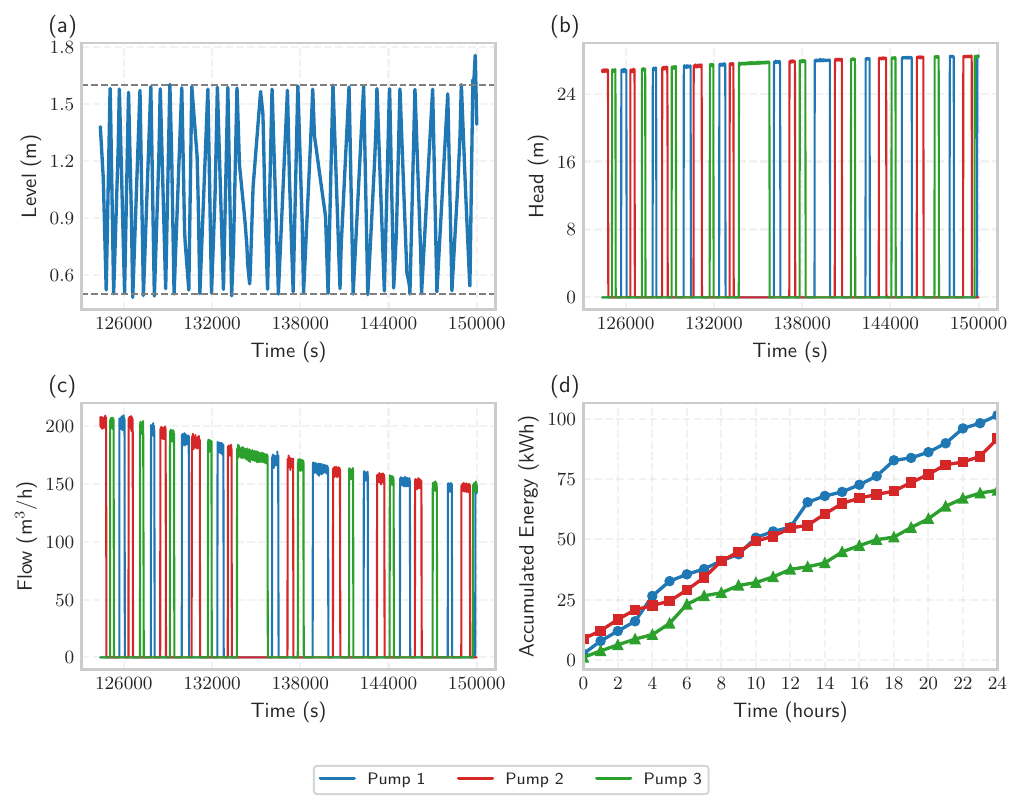}
    \caption{Pump and station metrics: (a) sump water level, (b) pump head, (c) pump flow rate, and (d) accumulated energy, considering system fault.}
    \vspace{-4mm}
    \label{fig:comsys}
\end{figure*}

There is a pronounced mismatch between supply and discharge, as shown in Fig.~\ref{fig:outtflow}. According to 
Fig.~\ref{fig:outtflow}-(a), the inflow rate, which is dominated by frequent, stochastic surges and fluctuates up to roughly $200~\mathrm{m^3/h}$. In contrast, Fig.~\ref{fig:outtflow}-(b) illustrates how the outflow steadily decreases from approximately $200~\mathrm{m^3/h}$ to around $150~\mathrm{m^3/h}$ over the given interval. This demonstrates how the increase in the friction coefficient due to pipe clogging reduces the station’s pumping capacity despite the unchanged control logic.
Fig.~\ref{fig:comsys} presents metrics including sump water level, head, flow rate, and accumulated energy. 

According to Fig.~\ref{fig:comsys}-(a), the absence of any excursions beyond 1.8~m in the level trajectory, despite a 25\% loss in outflow rate, demonstrates the controller’s resilience, as the high-level alarm and dual-pump operation prevent overflow.

\paragraph{System fault profile}
To simulate the increase in dynamic head loss, a ramp operator is defined as follows:
\[
r(t, \tau_0, \tau_1) = 
\begin{cases}
0, & t < \tau_0, \\
\frac{t - \tau_0}{\tau_1 - \tau_0}, & \tau_0 \leq t \leq \tau_1, \\
1, & t \geq \tau_1,
\end{cases}
\]
where $r \in [0, 1]$, and $\tau_0$ and $\tau_1$ represent the start and end times of the system fault, respectively. Using this ramp operator, the friction coefficient is updated affinely as follows:

\[
k(t) = k_{0} \left( 1 + \Delta k \, r(t; \tau_{0}, \tau_{1}) \right).
\]

Here, $k_0$ and $\Delta k$ denote the initial friction coefficient and the relative increment, respectively. If the formation of air pockets causes a change in static head, the change can be modeled as:
\[
H_{\text{static}}(t) = H_{\text{static},0} + \Delta H_{\text{static}} \, r(t; \tau_{0}, \tau_{1}),
\]
where $H_{\text{static},0}$ is the initial static head, and $\Delta H_{\text{static}}$ is the absolute increase in static head once the fault is fully developed.

\paragraph{system fault simulation results}
The simulation is carried out assuming that pipe clogging begins to form on the second day, within the interval $t \in [40000, 61600]$, which is equivalent to 5.5 hours. The initial friction coefficient is $k_0 = 0.0006$, with an increment of 100\%, resulting in a doubling of the friction by the end of the fault time interval. 

Moreover, the static head $H_{\text{static}_0}$ remains the same as before and is equal to 2 m, while the increase in static head is set to $H_{\text{static,inc}} = 0.5~\text{m}$. Similar to the pump-related scenario, key performance metrics are provided to evaluate the impact of the system fault on these variables. 

At fixed speed, a system fault steepens the system curve, shifting the operating point along the pump curve, with higher head and lower flow. These two implications are clearly observable in Fig.~\ref{fig:comsys}-(b) and Fig.~\ref{fig:comsys}-(c).

\begin{figure}
    \centering
    \includegraphics[width=\columnwidth]{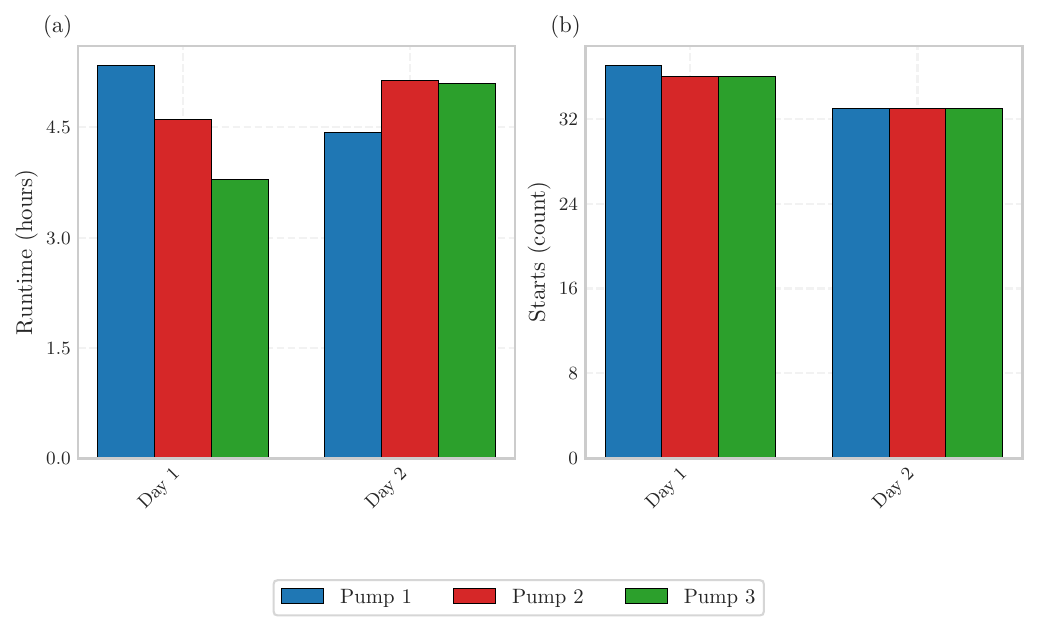}
    \caption{Runtime (a) and number of starts (b) for each pump over a two-day simulation interval, considering the system fault scenario.}
    \label{fig:metricsystem}
\end{figure}

The energy rise in Fig.~\ref{fig:comsys}(d) results from longer runtimes rather than power increase. Runtime and start count are shown in Fig.~\ref{fig:metricsystem}.

\begin{figure}
    \centering
    \includegraphics[width=\columnwidth]{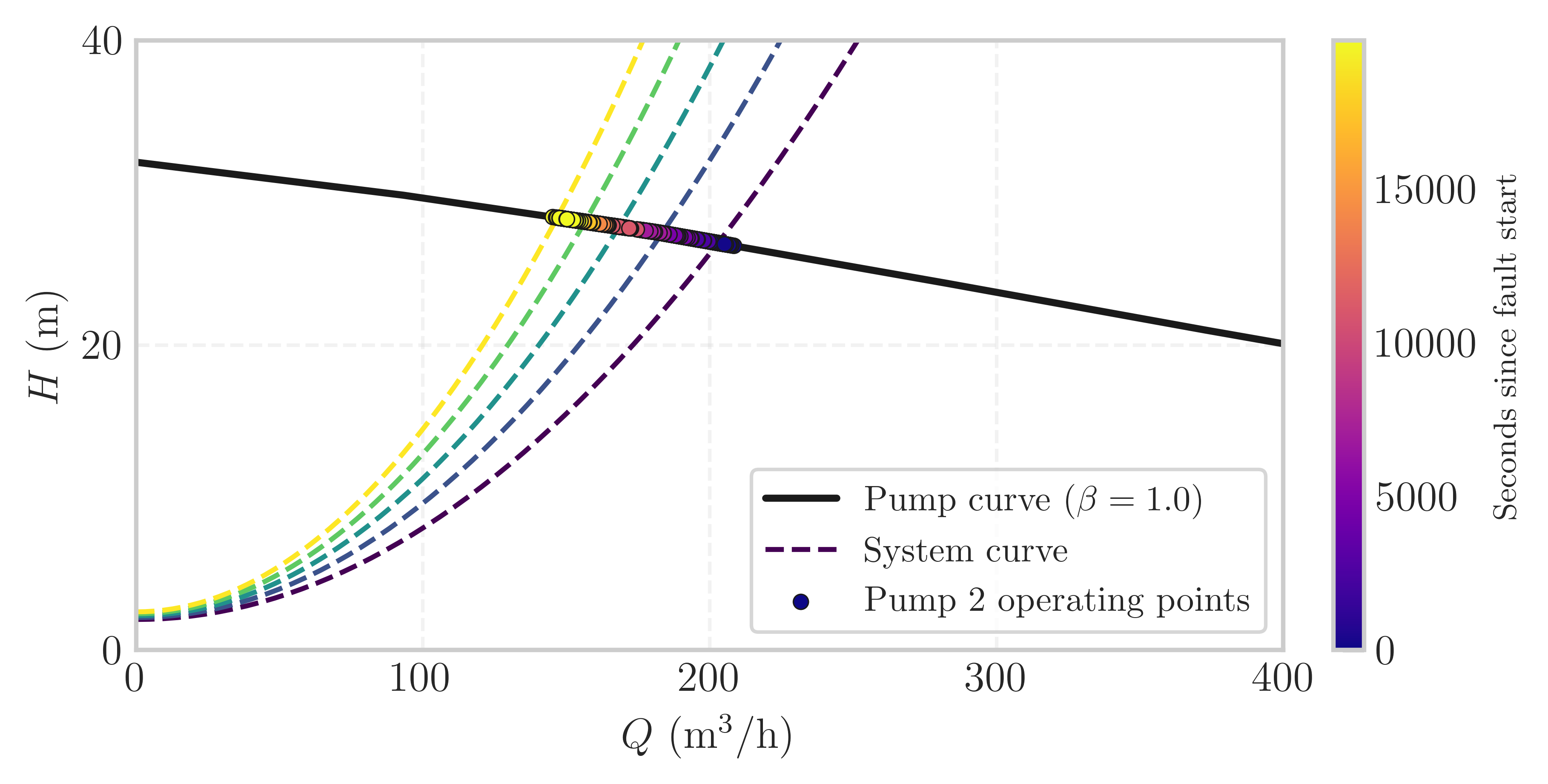}
    \caption{Trajectory of the operating point during the system fault scenario.}
    \vspace{-4mm}
    \label{fig:sysfault}
\end{figure}

As seen in Fig.~\ref{fig:metricsystem}(a), runtime increases on the second day except for Pump 1, as flow decreases. Cycle elongation reduces start opportunities as shown in Fig.~\ref{fig:metricsystem}(b), and the uniform reduction confirms the fault affects all pumps. The trajectory of the operating point during pipe clogging is depicted in Fig.~\ref{fig:sysfault}.

Consistent with theory, pipe clogging shifts the operating point leftward along the pump curve, into a lower-flow, higher-head region.
\section{Fault detection algorithm}
 \label{sec:dectection}
Corrective maintenance typically incurs significant costs. Misdiagnosing the fault origin leads to incorrect personnel dispatches, extra mobilizations, longer outages, and escalating unplanned work. A five-year internal maintenance report from SVOA for a municipal wastewater pump station shows that preventive maintenance constitutes only $4.4\%$ of total costs, planned corrective maintenance accounts for $61.2\%$, and unplanned corrective maintenance for $34.3\%$. Together, planned and unplanned corrective maintenance represent about $95.6\%$ of the total.


Unplanned corrective maintenance refers to emergency dispatches, whereas planned corrective maintenance represents failures that have already occurred but for which the maintenance dispatch is scheduled. Misdispatch of technician teams and spare parts commonly occurs in these types of maintenance tasks. This highlights that developing a reliable pump and system fault isolation methodology can yield multiple benefits. First, it ensures that the right technicians and spare parts, such as those required for pump removal or main flushing, are dispatched. This shortens response times and reduces travel costs, which represent a major portion of maintenance operations. Moreover, it helps convert emergent events into scheduled interventions. Hence, the isolation methods introduced here are not only diagnostically valuable but also economically justified.
\subsection{Problem definition}
As illustrated in Fig.~\ref{fig:operating_point}, the operating point of a wastewater pump in a pressurized network is defined by the intersection of two one-dimensional curves in the $(Q, H)$ plane, which are referred to as characteristic curves.

The pump curve, $H_{\text{pump}}$, is parameterized by both frequency and the health condition of the pump's hydraulic mechanism, represented as $a(t) = (a_0(t), a_1(t), a_2(t))^\top$. In contrast, the system curve, $H_{\text{system}}$, is determined by the static head and piping losses, expressed as $b(t) = (H_{\text{static}}(t), k(t))^\top$. Let:
\[
\Gamma_p(t) = \{(Q, H) \mid H = H_{\text{pump}}(Q; f(t), a(t))\},
\]
\[
\Gamma_s(t) = \{(Q, H) \mid H = H_{\text{system}}(Q; b(t))\}.
\]

Here, $\Gamma_p(t)$ and $\Gamma_s(t)$ denote the pump and system curves at time $t$, respectively. Thus, the operating point $(Q_t, H_t)$ is found by $(Q_t, H_t) = \Gamma_p(t) \cap \Gamma_s(t)$ and is subject to different behaviors under unhealthy operating conditions.

\subsubsection{Movement along the system curve}
If \( \Gamma_s \) remains constant while \( \Gamma_p \) drifts, then the operating point slides along the system curve:
\[
(Q_t, H_t) \in \Gamma_s(t), \quad \text{with} \quad \dot{b}(t) = 0, \ \dot{a}(t) \neq 0,
\]
where $(\dot{\;})$ denotes the time derivative. This behavior may arise due to the VFD mechanism or degradation in the pump itself over time. As shown in Fig.~\ref{fig:trajectory}, the trajectory of \( (Q_t, H_t) \) is constrained to \( \Gamma_s(t) \). If the pump curve is already scaled by affinity laws to eliminate the effect of frequency variation, and at least one \( \dot{a}(t) \neq 0 \) still holds, then the movement of the operating point can be attributed to pump degradation.

\subsubsection{Movement along the pump curve}
If the pump characteristics remain constant but the system undergoes changes such as valve throttling, pipe clogging, or similar disturbances, then the operating point is expected to move along a single pump curve, as illustrated in Fig.~\ref{fig:sysfault}:
\[
(Q_t, H_t) \in \Gamma_p(t) \quad \text{with} \quad \dot{a}(t) = 0, \ \dot{b}(t) \neq 0.
\]

Therefore, if time series data is recorded as $(Q_t, H_t, f_t)$, the trajectory is defined as $\mathcal{C} = \{(Q_i, H_i)\}_{i=0}^{T-1} \subset \mathbb{R}^2$. The geometric behavior of the trajectory \( \mathcal{C} \) helps distinguish system-side changes from pump-side effects. In other words, the path traced by the operating point acts as a low-dimensional signature that can be used as a reliable indicator for fault detection frameworks.

Considering the presence of noise and limited sampling, reliably distinguishing the movement of the operating point can be challenging. To address this, two approaches are introduced in the following subsections. First, a statistical nested model is proposed to support informed decision-making. Second, a tangent residual method provides a mathematical framework to distinguish pump-side from system-side faults.
\subsection{Statistical nested F-test method}
\label{subsec:ftest}
To diagnose pump degradation, a hypothesis test is formulated to compare two competing models:
\[
\begin{cases}
\mathcal{H}_0 : \text{Pump parameters stay constant} \\
\mathcal{H}_1 : \text{Pump degrades over time}
\end{cases}
\]

A fair null model $\mathcal{H}_0$ is established by assuming that the pump curve coefficients are unknown constants, which are estimated from the data rather than fixed at nominal values \cite{DraperSmith1998}. In other words, $(Q_t, H_t, f_t)$ are observed for $t = 0, \ldots, T-1$, and the null model is expressed as
\[
H_t=H_{\mathrm{pump}}(Q_t,f_t;\theta_t)+\varepsilon_t,\qquad 
\varepsilon_t\stackrel{\text{i.i.d.}}{\sim}\mathcal N(0,\sigma^2).
\]

The two nested models are then compared:
\begin{align*}
\mathcal{H}_0 &: \theta_t \equiv \theta = (a_0, a_1, a_2)^\top, \\
\mathcal{H}_1 &: \theta_t = (a_0 + \alpha_0 t, \ a_1 + \alpha_1 t, \ a_2 + \alpha_2 t)^\top.
\end{align*}

The null model has three free parameters, denoted $p_0$, and assumes time-invariant coefficients. 
The alternative model incorporates a linear time drift and is nested within the null model. 
Setting slope parameters $\alpha$ to zero reduces the degradation model to the null. The alternative has six free parameters $p_1$. The pump curve is already scaled by affinity laws to account for the VFD. After normalization, pump head depends on a few interpretable coefficients, such as the shut-off and slope terms. Pump failures and gradual hydraulic wear manifest as a drift in these coefficients, whereas a healthy pump keeps them constant over time.

The sum of squared residuals (SSR) is computed for each model $j$ as follows, with ordinary least squares optimization used to estimate parameters \cite{BatesWatts1988}:
\[
SSR_j(\boldsymbol{\theta}) = \sum_{t=0}^{T-1} \left[ H_t - H_{\text{pump}}(Q_t, f_t, \boldsymbol{\theta}_j) \right]^2, \quad j \in \{0, 1\}.
\]

Since the two models are nested, the lack of fit can be quantified using an F-test:
\[
F = \frac{(SSR_0 - SSR_1)/(p_1 - p_0)}{SSR_1 / (m - p_1)},
\]
where $m$ is the number of data points, $p_0$ and $p_1$ are the number of free parameters in models $\mathcal{H}_0$ and $\mathcal{H}_1$, respectively. Under the null hypothesis $\mathcal{H}_0$, the F-statistic follows an F-distribution:
\[
F \sim F_{p_1 - p_0, \, m - p_1}
\]

In addition, the Akaike Information Criterion (AIC) is used to assess the trade-off between goodness of fit and model complexity \cite{Akaike1974}. It is formulated as:
\[
AIC = m \ln\left(\frac{SSR}{m} \right) + 2p,
\]
where $p$ is the number of model parameters. As an illustrative example, simulated data for $m = 50$ time steps is generated. Additive degradation is introduced into the data generator as follows:
\[
\begin{aligned}
A_{0,t} &= 15 - 0.1\,t, \\
A_{1,t} &= 5\times 10^{-4} + 10^{-6}\,t, \\
A_{2,t} &= 9\times 10^{-4} + 5\times10^{-6}\,t.
\end{aligned}
\]

Linear degradations are chosen to generate a moderate, early-stage loss of head over a 50-step horizon. 
This loss is large enough to be statistically detectable under measurement noise, yet small enough to remain within a realistic pre-failure domain. 
At a representative flow $Q$, the head drop is
\[
\Delta H \approx \Delta a_{0} -  \, \Delta a_{1}Q \, - \Delta a_{2}Q^{2}.
\]

For the representative flow rate of $30~\frac{\text{m}^3}{\text{s}}$, the induced drop over the entire time interval equals $10.25 \,\text{m}$. 
This corresponds to a signal change several standard deviations above the injected noise, thus creating a nontrivial but solvable detection task. 

At each time step, the operating point is computed, where in this example it is assumed that
\[
H_{\text{system}} = 2.0 + 6\times 10^{-4} Q^2.
\]

Note that this setup does not correspond to a real network; the synthetic $H_{\text{pump}}$ and $H_{\text{system}}$ 
are chosen solely to validate the statistical approach while maintaining a realistic operating region for the study. 
Additionally, zero-mean Gaussian noise is added to simulate measurement uncertainty, 
with $\sigma_Q = 1 \,\text{m}^3/\text{h}$ and $\sigma_H = 0.5 \,\text{m}$.

The pump's operating frequency starts at 50 rpm and gradually decays over time:
\[
f_t = 50 - 0.1t + \varepsilon_t, \quad \varepsilon_t \sim \mathcal{N}(0, 1).
\]

\begin{figure}
    \centering
    \includegraphics[width=0.9\columnwidth]{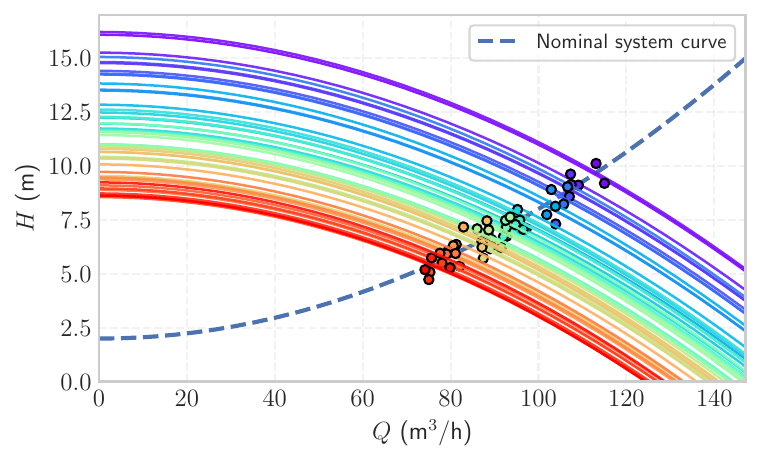}
    \caption{Operating point shift due to progressive degradation with noise.}
    \vspace{-4mm}
    \label{fig:degradation}

\end{figure}

This drift moves the operating speed by approximately 5 units over 50 steps, with a jitter of about 1 unit. 
The affinity laws are used to normalize the pump curve, ensuring that the degradation model focuses solely on the degradation phenomenon itself. 
Therefore, the inclusion of a mild exogenous speed ramp serves to stress-test the affinity law normalization. The distribution of simulation points is shown in Fig.~\ref{fig:degradation} and the statistical outcomes are summarized in Table 1.

\begin{table}[!t]
\caption{Statistical results.\label{table:statisticalresults}}
\centering
\begin{tabular}{|c||c|}
\hline
\textbf{Statistic} & \textbf{Value} \\
\hline
F-ratio & 196.58 \\
\hline
p-value & $1.11 \times 10^{-16}$ \\
\hline
AIC & $\text{AIC}_0 = 89.31, \quad \text{AIC}_1 = -38.07$ \\
\hline
\end{tabular}
\vspace{-4mm}
\end{table}

\begin{figure}
    \centering
    \includegraphics[width=0.9\columnwidth]{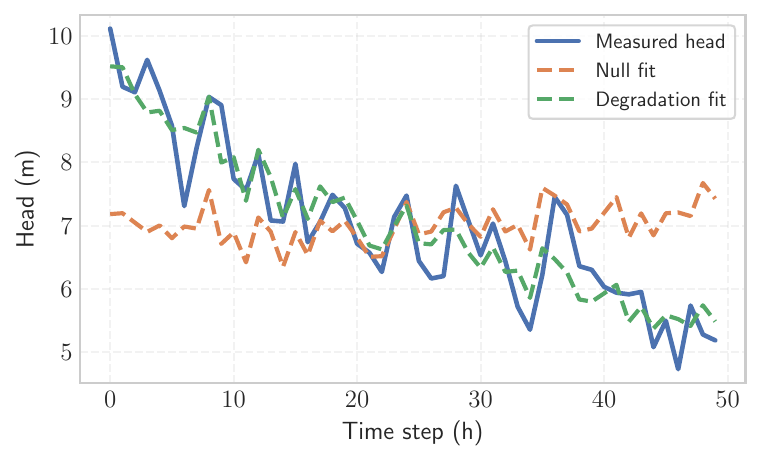}
    \caption{Comparison of measured pump head over time with two model fits: a null model and a degradation model.}
    \vspace{-4mm}
    \label{fig:fitting}
\end{figure}

Based on the results in Table 1, both the F-test and the AIC strongly support the degradation model. The degradation model provides a superior explanation of the data. The fitted trajectories of the noisy measured heads under the hypotheses  $\mathcal{H}_0$ and $\mathcal{H}_1$ are visualized in Fig.~\ref{fig:fitting}.

As Fig.~\ref{fig:fitting} illustrates, the null model fit oscillates around 22 meters and fails to capture the underlying degradation trend. In contrast, the alternative hypothesis closely aligns with the numerical results, accurately reflecting the system's deterioration over time.

This analysis demonstrates that statistically significant degradation can be detected before the head reaches a critical threshold, enabling timely maintenance and intervention.

\subsection{Tangent residual method}
The mathematical approach to decoupling pump degradation from system faults is to derive the differential test. As previously discussed, to eliminate pure speed effects such as soft start, soft stop, and setpoint changes, each observation at frequency $f(t)$ is mapped to a fixed nominal frequency $f_{\text{nominal}}$ using the affinity laws:
\[
Q_t^{\star} = Q(t) \frac{f_{\text{nominal}}}{f(t)}, 
\quad
H_t^{\star} = H(t) \left( \frac{f_{\text{nominal}}}{f(t)} \right)^{2}.
\]
Thereby, a family of pump curves $H_{\text{pump}}(\cdot, N)$ collapses into a single nominal-speed pump curve, expressed as follows:
\[
\tilde{H}_{\text{pump}}(Q) = H_{\text{pump}}(Q, N = 1) = a_{1} + a_{2}Q + a_{3}Q^{2}.
\]
While the system curve remains unchanged, as in \eqref{eq:systemcurve}, the operating point $(Q_t^{\star}, H_t^{\star})$ after normalization is given by:

\begin{equation}
H_t^{\star} = \tilde{H}_{\text{pump}}(Q_t^{\star}; a_t) = H_{\text{system}}(Q_t^{\star}; b_t). 
\label{eq:equilibrium}
\end{equation}

The local slopes at the operating point can be determined using partial derivatives with respect to the flow rate. In other words,

\begin{equation}
\begin{aligned}
m_p(t) &:= \frac{\partial \tilde{H}_{\text{pump}}(Q_t^{\star}; a_t)}{\partial Q} 
= a_{2} + 2 a_{3} Q_t^{\star}, \\
m_s(t) &:= \frac{\partial H_{\text{system}}(Q_t^{\star}; b_t)}{\partial Q} 
= 2 k Q_t^{\star}.
\end{aligned}
\label{eq:local_slopes}
\end{equation}

Here, $m_p(t)$ and $m_s(t)$ are the local slopes of the pump curve and system curve at the operating point, respectively. By differentiating the equilibrium \eqref{eq:equilibrium}, the crucial balance term is obtained:

\begin{equation}
\left( m_s(t) - m_p(t) \right) \dot{Q}_t^{\star} = \underbrace{\frac{\partial \tilde{H}_{\text{pump}}(Q_t^{\star}; a_t)}{\partial a}}_{\text{pump drift}} \dot{a}_t - \underbrace{\frac{\partial H_{\text{system}}(Q_t^{\star}; b_t)}{\partial b}}_{\text{system drift}} \dot{b}_t.
\label{eq:identification}
\end{equation}
Equivalently, the tangent contribution is subtracted from $H_t^{\star}$ in two ways:
\begin{equation}
\begin{aligned}
\Psi_s(t) := \dot{H}_t^{\star} - m_s(t)\,\dot{Q}_t^{\star}
= \frac{\partial H_{\text{system}}(Q_t^{\star}; b_t)}{\partial b} \cdot \dot{b}_t, \\
\Psi_p(t) := \dot{H}_t^{\star} - m_p(t)\,\dot{Q}_t^{\star}
= \frac{\partial \tilde{H}_{\text{pump}}(Q_t^{\star}; a_t)}{\partial a} \cdot \dot{a}_t.
\end{aligned}
\label{eq:detection}
\end{equation}

Based on \eqref{eq:detection}, if the system is fixed (i.e., $\dot{b}_t = 0$), then $\Psi_s(t) = 0$, and the velocity vector $(\dot{Q}_t^{\star}, \dot{H}_t^{\star})$ is tangent to the system curve. In other words, the operating point moves along the system curve because the pump has undergone changes such as degradation.

On the other hand, if the pump remains unchanged (i.e., $\dot{a}_t = 0$), then $\Psi_p(t) = 0$, and the motion is along the pump curve while the system has changed due to factors such as valve throttling, clogging, or similar effects.

Moreover, a decision index can be defined using \eqref{eq:detection}:
\[
\mathcal{I}(t) = \frac{|\Psi_p(t)|}{|\Psi_p(t)| + |\Psi_s(t)|} \in [0,1].
\]

Here, the index $\mathcal{I}(t)$ can be interpreted as the continuous attribution of operating-point movement to pump faults versus system faults. Over a time window $W$, the index is calculated as:
\[
\mathcal{I}_W = \frac{\mathbb{E}_W \left[ |\Psi_p| \right]}{\mathbb{E}_W \left[ |\Psi_p| \right] + \mathbb{E}_W \left[ |\Psi_s| \right]}.
\]
where, if $\mathcal{I}_W \approx 0$, there is a system change, and if $\mathcal{I}_W \approx 1$, the pump has deteriorated. By utilizing a block bootstrap \cite{Künsch1989, Lahiri2003}, a $100(1 - \alpha)\%$ confidence interval (CI) is estimated for $\mathcal{I}_W$. Subsequently, if $\mathrm{LCI}(\mathcal{I}_W)$ greater than 0.6, the change is labeled as a pump fault, and if $\mathrm{UCI}(\mathcal{I}_W)$ less than 0.4, the change can be attributed to the system. The choice of $\alpha$ affects the width of the confidence interval and the reliability of fault detection. A smaller $\alpha$ results in a wider interval, reducing the number of false alarms but potentially increasing missed detections, 
whereas a larger $\alpha$ narrows the interval and increases sensitivity to faults at the cost of higher false-positive rates.

The illustrative example in subsection~\ref{subsec:ftest} is used again to empirically validate the tangent residual method 
for fault discrimination. First, each observed operating point is normalized to the pump's nominal rotational speed, 
yielding the transformed coordinates $(Q_t^{\star}, H_t^{\star})$ and isolating the effect of pump failure from speed control actions. 
The decision index $\mathcal{I}_W(t)$ is computed at each time step to quantify the instantaneous distinguishability of the operating point movement. 
The window-averaged index $\mathcal{I}_W(t)$ over the entire 50 time steps equals 0.746.

\begin{figure}
    \centering    \includegraphics[width=0.9\columnwidth]{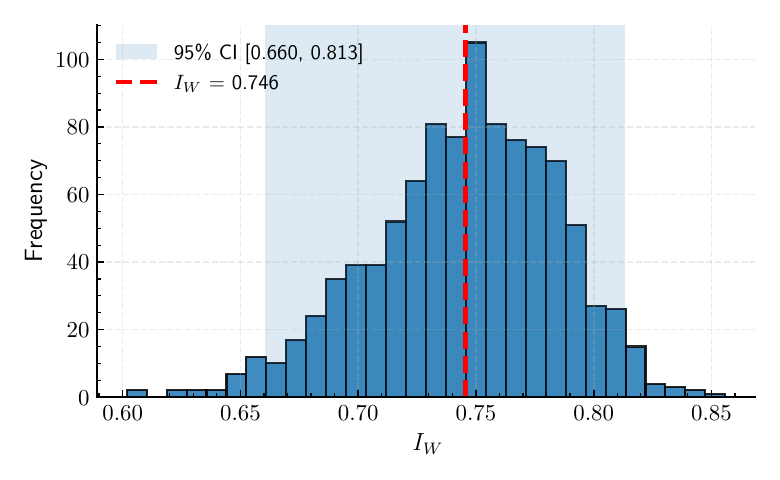}
    \caption{Bootstrap distribution of the index $\mathcal{I}_W$.}
    \vspace{-4mm}
    \label{fig:Iw}
\end{figure}

To assess the statistical significance of this result, a block bootstrap procedure with a block size of five hours 
is applied to estimate a 95\% confidence interval for $\mathcal{I}_W$, which is determined to be $[0.660, \, 0.813]$. 
The bootstrap distribution of $\mathcal{I}_W(t)$ is illustrated in Fig.~\ref{fig:Iw}. 
As the entire confidence interval lies substantially above the decision threshold of 0.6, 
the result provides statistically robust evidence that the observed performance originates from the pump side.

\begin{figure}
    \centering    \includegraphics[width=0.9\columnwidth]{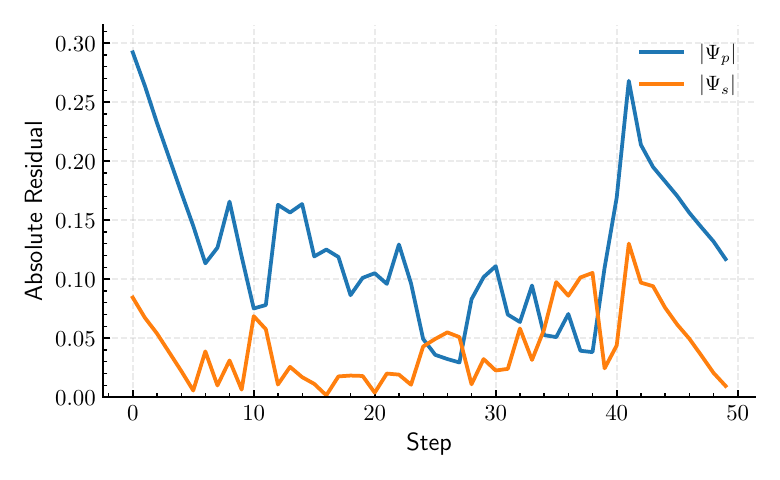}
    \caption{Evolution of the absolute pump and system residuals over time steps.}
    \vspace{-4mm}
    \label{fig:tangentresidual}
\end{figure}

Furthermore, the time series of tangent residuals in Fig.~\ref{fig:tangentresidual} consistently shows greater magnitude in $\Psi_p$ compared to $\Psi_s$, 
corroborating the dynamics of equilibrium shifts caused by changes in the pump's hydraulic parameters. 
Therefore, the method correctly identifies pump degradation despite the presence of confounding factors 
such as speed variations and measurement noise.

It should be noted that thresholds are not absolute constants; in this part, they were selected based on the governing physics of the example to provide a conservative diagnostic margin. The optimal thresholds may depend on pump and system characteristics as well as sensor noise profiles. Therefore, operational deployment entails a data-driven calibration phase using historical fault data to establish adaptive thresholds that continue to learn from ongoing operational data. This adaptive thresholding mechanism is developed and implemented in the following section.

\section{Numerical results} \label{sec:numerical}
This section applies the fault-origin discriminators developed in Section~\ref{sec:dectection} and evaluates their performance on simulation data. Subsection~\ref{subsec:dataset} introduces the dataset and the preprocessing and normalization steps forming the basis for evaluation. The nested-model F-test and the tangent residual method are applied to data covering normal operation, pump faults, and system faults, with their capabilities compared through a confusion matrix. The nested-model F-test is further extended to distinguish among these three states. Subsection~\ref{subsec:methods} presents the detection results and evaluates how reliably the methods flag faults, localize their onsets, and distinguish between pump and system faults using only three SCADA measurements. Finally, it discusses their applicability as online detectors in real-world setups.

\subsection{Dataset}\label{subsec:dataset}
The dataset comprises simulation results for the fault scenarios in Subsections~\ref{subsub:blockagge} and \ref{subsub:clogging}, with identical failure functions, onset times, and durations. It represents two days of pump station operation with three pumps: Pump~1 experiences an internal blockage, and a system fault affects the entire station on the second day.

For diagnosis, only three pump metrics are used: flow rate, head, and frequency. Since Pump~1 is blocked, its metrics are primarily analyzed. The data resolution is 1~s, and because the probability of a fault within one second is negligible, a time window is defined for analysis.

True labels (normal, pump fault, system fault) are available for each time window but used only for performance assessment. The diagnostic methods can operate online without knowledge of the labels. Finally, the dataset is filtered to points near the nominal frequency, and affinity laws are applied to scale flow and head for consistent comparison.
\subsection{Simulation results} 
\label{subsec:methods}
In contrast to the illustrative example, it is assumed here that the pump curve is not known. 
A quadratic model is fitted using fault-free data to parameterize the nominal curve. 
The nested F-test is then applied to each operating cycle of Pump~1. 
The null hypotheses for both the pump and system curves are based on the nominal conditions, 
whereas the alternative hypotheses assume time-varying coefficients. 
For every cycle, the F-statistic and p-value are computed.

In the tangent residual method, 25 samples are grouped into analysis segments to define the time window for online decisions and to exclude non-informative transients. 
The classification thresholds are not fixed. 
Instead, the tangent residual detector runs a learning phase during the first six hours of the dataset 
to establish baseline behavior. 
The thresholds are then periodically updated using the most recent normal data. 
All valid segments are processed and compared against known fault windows for validation.

\begin{figure}
    \centering    \includegraphics[width=0.9\columnwidth]{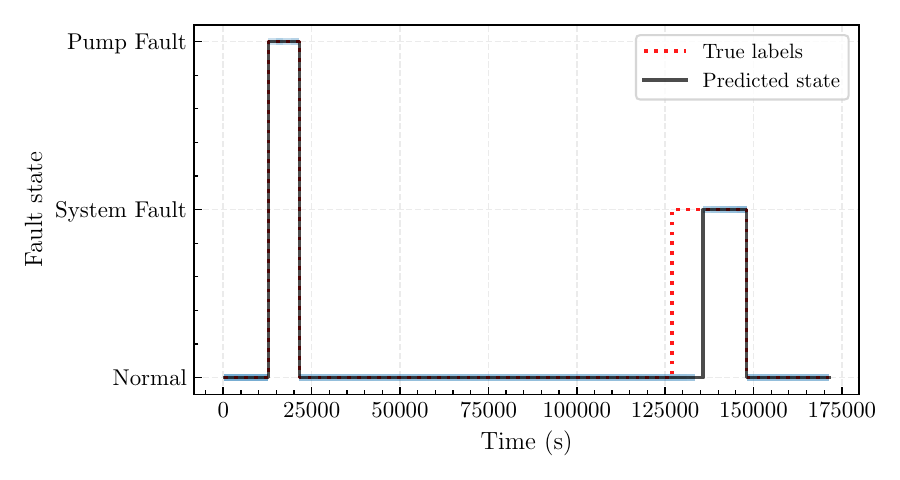}
    \caption{Estimated fault states over time via the tangent residual method.}
    \label{fig:fault_states}
\end{figure}

State classification over time is shown in Fig.~\ref{fig:fault_states}, which illustrates that the predicted states closely follow the true labels with no confusion between pump and system faults. 
A slight delay in detecting the system fault is observed, which may be due to its more subtle signatures in the data. 

\begin{figure}
    \centering    \includegraphics[width=0.9\columnwidth]{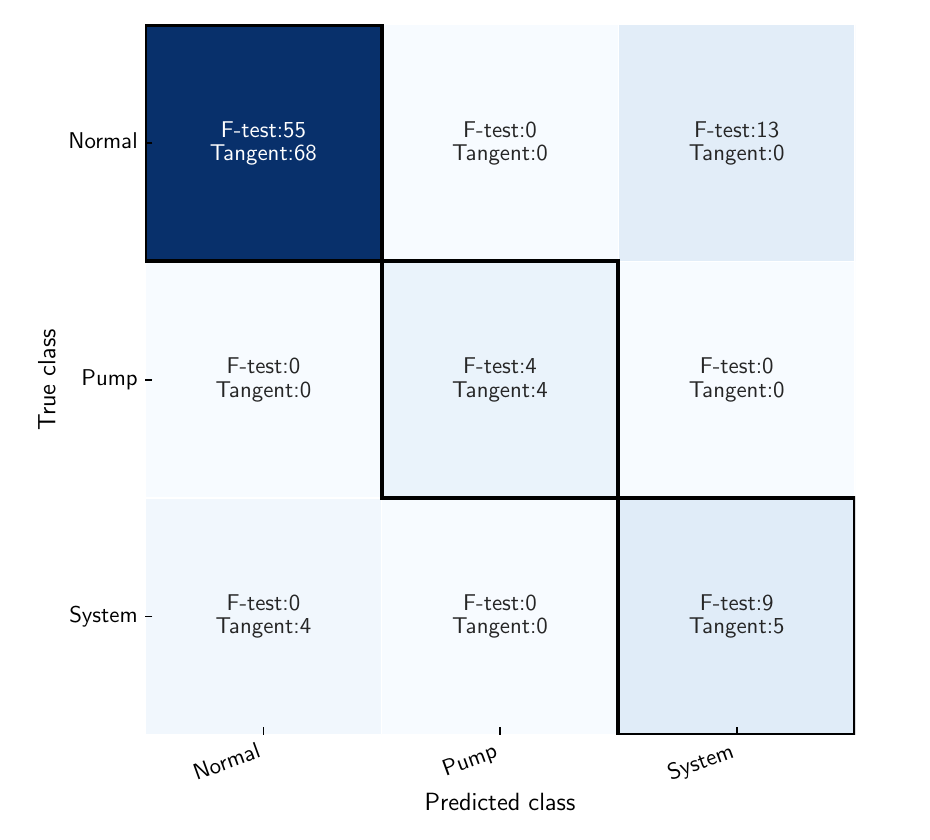}
    \caption{Confusion matrix of the F-test and tangent residual methods, with each cell indicating the number of samples.}
    \vspace{-4mm}
    \label{fig:cm}
\end{figure}

Moreover, the confusion matrix of the two methods, based on the classification of 81 cycles, is presented in Fig.~\ref{fig:cm}. The resulting confusion matrix metrics for these methods are listed in Table~\ref{table:comparison}. 

\begin{table}[!t]
\caption{Comparison of diagnostic methods.\label{table:comparison}}
\centering
\setlength{\tabcolsep}{4pt}
\begin{tabular}{|c||c|c|c|}
\hline
\textbf{Method} & \textbf{Avg. precision} & \textbf{Avg. recall} & \textbf{Avg. F1-score} \\
\hline
Nested F-test & 0.80 & 0.94 & 0.82 \\
\hline
Tangent residual & 0.981 & 0.852 & 0.895 \\
\hline
\end{tabular}
\vspace{-4mm}
\end{table}

The confusion matrix shows that both methods correctly distinguish normal operation, pump faults, and system faults. Table~\ref{table:comparison} indicates that although the nested F-test achieves higher recall, it suffers from lower precision, resulting in more false positives. In contrast, the tangent residual method attains markedly higher precision and a superior overall F-score, highlighting its diagnostic reliability.
\section{Conclusion}
\label{sec:conclusion}
This paper presented a physics-enhanced, parameter-driven simulator for three-pump wastewater stations and a framework for fault detection and origin isolation. The simulator reproduces transient hydro-electro-mechanical behavior at one-second resolution, supports station adaptation via physical parameters, and generates faulty datasets when real failures are scarce. Two diagnostic methods, namely the nested-model F-test and the tangent residual index, were proposed to distinguish pump failures from system failures using only flow, head, and frequency. Together, these form a toolbox for what-if studies, early diagnosis, and condition-based maintenance.  

Validation against high-frequency SCADA data from a municipal station showed strong fidelity across key metrics. The simulator captured water-level dynamics with a normalized mean absolute error of 7.48\% and reproduced daily start counts and runtimes for all three pumps with negligible deviations.

On the diagnostic side, the proposed methods successfully separated internal pump blockage from pipe clogging and delivered reliable, quantifiable indicators for maintenance decisions. Across 81 simulated operating cycles, they achieved promising confusion-matrix results, indicating accurate and interpretable fault-origin isolation suitable for online use and enabling timely, targeted interventions.  

This paper analyzed open-loop control with soft-start/stop mechanisms, which is the most widely used approach in municipal wastewater pump stations. Future work should integrate closed-loop level control and transient hydraulic phenomena such as water hammer and entrained air. Extending the fault set and coupling diagnostics with cost-aware maintenance scheduling remain important directions for research.

In addition to wastewater applications, the proposed diagnostic framework can be generalized to other coupled systems, including district heating systems, HVAC plants, and similar infrastructures, where the operating point is likewise determined by the intersection and shifts predictably under component degradation or network changes. Since the fault isolation framework relies on routine measurements and residual analysis, it can be adapted to detect faults in any process governed by similar principles.

\section*{Acknowledgments}
We gratefully acknowledge the collaboration of Stockholm Vatten och Avfall (SVOA) and the City of Stockholm, whose operational context and valuable discussions have informed this work. This research was supported by Digital Futures, a cross-disciplinary research center at KTH, Sweden.

\bibliographystyle{IEEEtran}
\bibliography{Biblio}


 



\vspace{-33pt}
\begin{IEEEbiography}[{\includegraphics[width=1in,height=1.25in,clip,keepaspectratio]{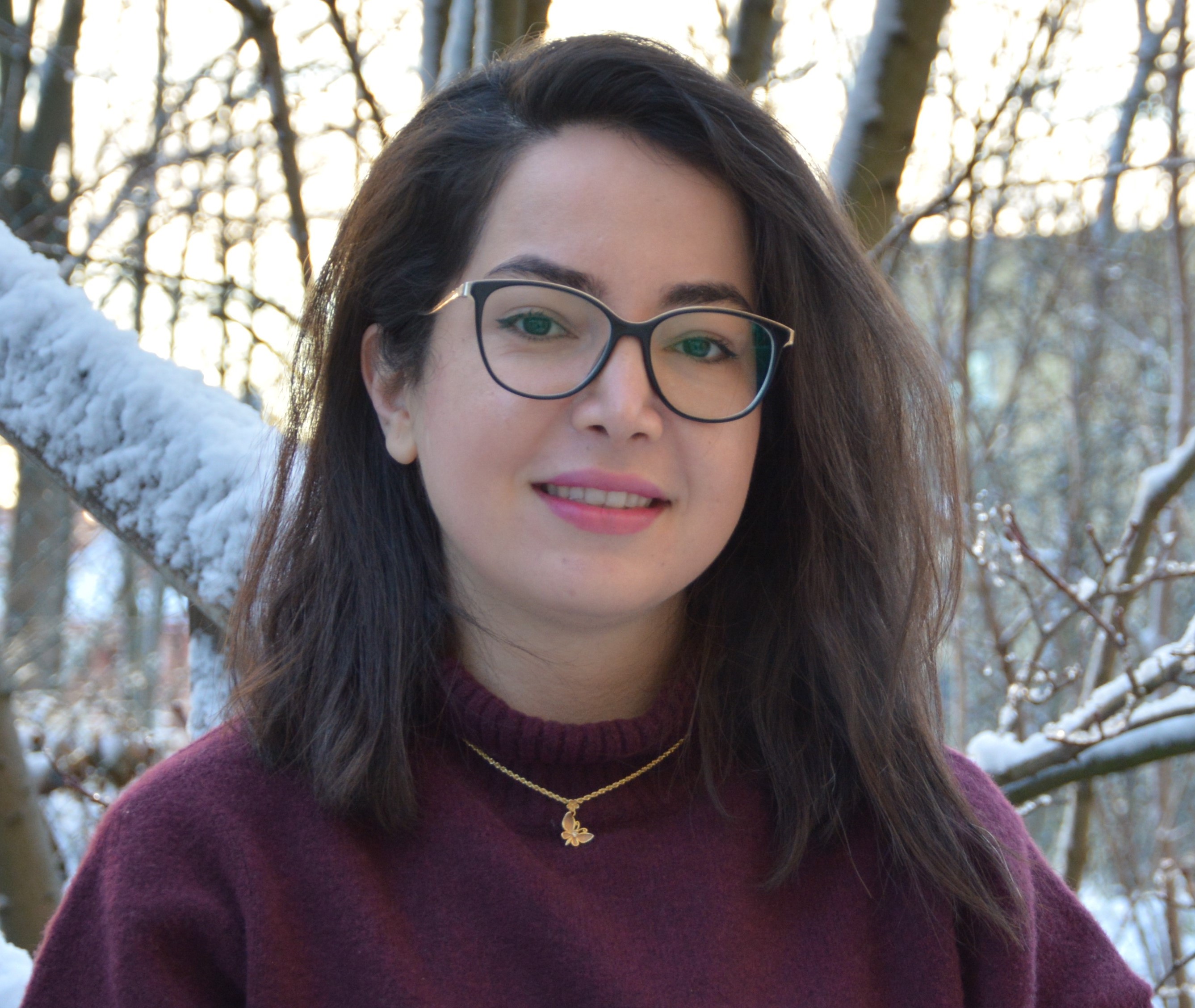}}]{Katayoun Eshkofti}
received her M.Sc. and Ph.D. degrees in Industrial Engineering from Amirkabir University of Technology (Tehran Polytechnic) and Ferdowsi University of Mashhad, Iran, respectively. Since 2024, she has been a postdoctoral fellow with the Division of Decision and Control Systems at KTH Royal Institute of Technology in Stockholm, Sweden. Her research interests include scientific machine learning, modeling, and simulation.
\end{IEEEbiography}
\vspace{-33pt}

\begin{IEEEbiography}[{\includegraphics[width=1in,height=1.25in,clip,keepaspectratio]{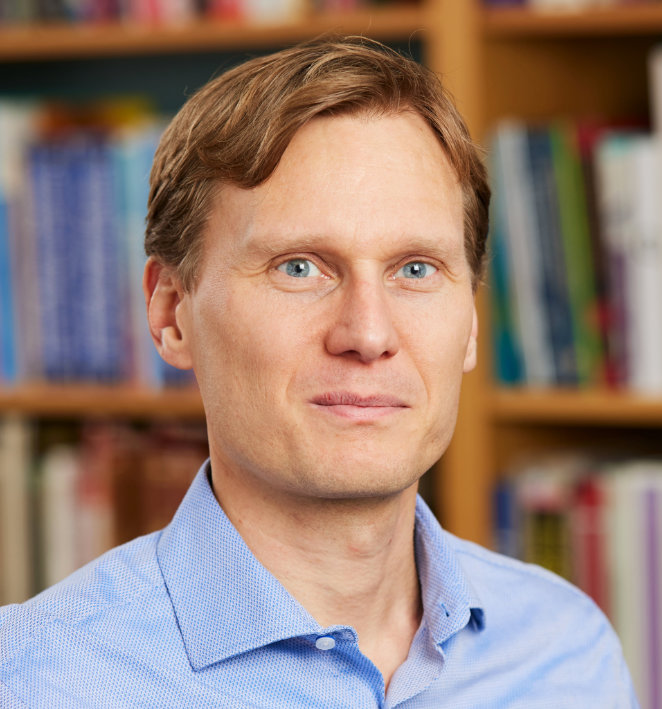}}]{Henrik Sandberg}
is Professor at the Division of Decision and Control Systems, KTH Royal Institute of Technology, Stockholm, Sweden. He received the M.Sc. degree in engineering physics and the Ph.D. degree in automatic control from Lund University, Lund, Sweden, in 1999 and 2004, respectively. From 2005 to 2007, he was a Postdoctoral Scholar at the California Institute of Technology, Pasadena, USA. In 2013, he was a Visiting Scholar at the Laboratory for Information and Decision Systems (LIDS) at MIT, Cambridge, USA. He has also held visiting appointments at the Australian National University and the University of Melbourne, Australia. His current research interests include security of cyber-physical systems, power systems, model reduction, and fundamental limitations in control. Dr. Sandberg was a recipient of the Best Student Paper Award from the IEEE Conference on Decision and Control in 2004, an Ingvar Carlsson Award from the Swedish Foundation for Strategic Research in 2007, and a Consolidator Grant from the Swedish Research Council in 2016. He has served on the editorial boards of IEEE Transactions on Automatic Control and the IFAC Journal Automatica, and is currently an elected member of the IEEE Control Systems Society Board of Governors. He is Fellow of the IEEE.
\end{IEEEbiography}

\begin{IEEEbiography}{Mikael Nilsson}

\end{IEEEbiography}

\begin{IEEEbiography}[{\includegraphics[width=1in,height=1.25in,clip,keepaspectratio]{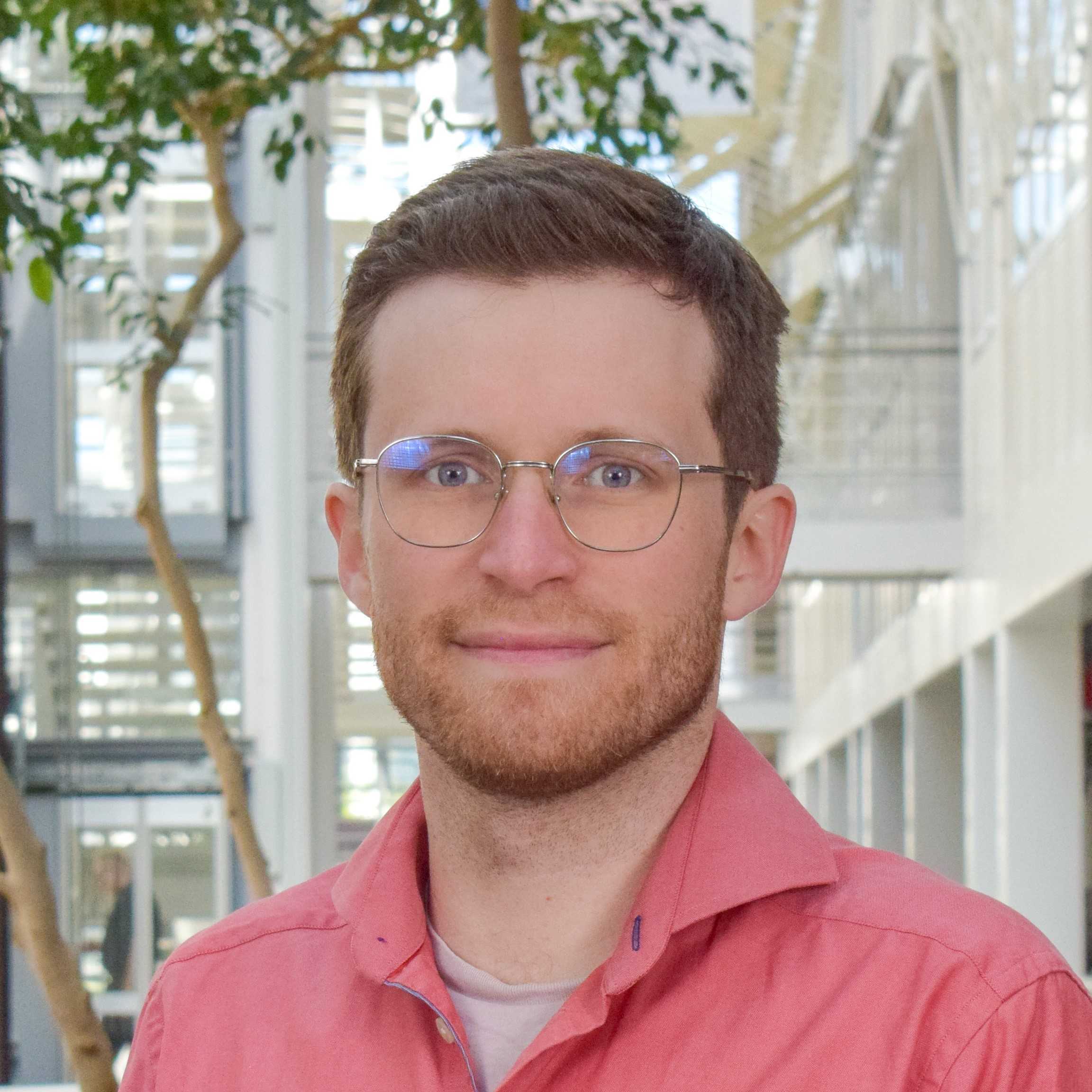}}]{Matthieu Barreau}
is an Assistant Professor in the Division of Decision and Control Systems at KTH Royal Institute of Technology, Stockholm, where he has been a faculty member since September 2023. He received his Ph.D. in Control Systems from LAAS-CNRS, Toulouse, in 2019, where his research focused on the stability analysis of coupled ordinary differential systems with string equations. His current research interests include physics-informed machine learning, traffic flow theory, infinite-dimensional systems, and controller synthesis.
He also holds a Master’s degree in Space Engineering from KTH Royal Institute of Technology, completed in 2016, and an Engineering degree in Aeronautical Engineering from ISAE-ENSICA, Toulouse, obtained in 2016. Prior to his current position, he served as an R\&D Manager at Tobii AB and as a Postdoctoral Researcher at KTH.
He has been recognized for his contributions to the field, notably receiving the Best French Ph.D. thesis award from GdR MACS and Club EEA in 2020.
\end{IEEEbiography}

\vfill

\end{document}